This is the author's accepted manuscript of the following article:

> Jeffrey Tsang, "The parametrized probabilistic finite-state transducer probe game player fingerprint model". *IEEE Transactions on Computational Intelligence and AI in Games* 2(3):208–224, September 2010.

The version of record of the article can be found at:

# The Parametrized Probabilistic Finite State Transducer Probe Game Player Fingerprint Model

Jeffrey Tsang, *Student Member, IEEE*

***Abstract*—Fingerprinting operators generate functional signatures of game players and are useful for their automated analysis independent of representation or encoding. The theory for a fingerprinting operator which returns the length-weighted probability of a given move-pair occurring from playing the investigated agent against a general parametrized probabilistic finite state transducer (PFT) is developed, applicable to arbitrary iterated games. Results for the distinguishing power of the 1-state opponent model, uniform approximability of fingerprints of arbitrary players, analyticity and Lipschitz continuity of fingerprints for logically possible players, and equicontinuity of the fingerprints of bounded-state probabilistic transducers are derived. Algorithms for the efficient computation of special instances are given; the shortcomings of a previous model, strictly generalized here from a simple projection of the new model, are explained in terms of regularity condition violations, and the extra power and functional niceness of the new fingerprints demonstrated. The 2-state deterministic finite state transducers (DFTs) are fingerprinted and pairwise distances computed; using this the structure of DFTs in strategy space is elucidated.**

## I. Introduction

GAME theory is the study of the interaction of multiple agents via their selection from a palette of moves. A simple, easily understood model, games are one of the most common tools in understanding any sort of interaction, whether sociological or psychological [1], economical [2], to biological [3]. Approaches vary from mathematical concepts such as Nash equilibria [4], computational methods such as population dynamics simulations [5], down to experimental testing of hypotheses drawn from analysis [6].

As an example, "Prisoner's Dilemma", one of the most famous two-move games, is used to model a wide variety of non-cooperative behaviors, finding applications as diverse as viral evolution [7], entrepreneur-venture capitalist relations [8] and international politics [9]. More generally, the payoff matrix for a two-move game can be varied, and the different sets of inequalities it satisfies govern which name it receives: another one is "Chicken", also considered in [9].

The author was with the Department of Mathematics and Statistics, University of Guelph, Guelph, ON N1G2W1 Canada. He is now with the Department of Computational Biology, University of Pittsburgh, Pittsburgh, PA 15260 USA (e-mail: jeffrey.tsang@ieee.org).

Iterated games, where the same choice is given to players multiple times, give rise to possibilities for responses to opponent's strategies and practically unlimited complexities. This lends itself naturally to the use of evolutionary algorithms for simulating populations of mutually interacting agents, varying all aspects of the conditions: this is the field of evolutionary game theory [10]. From there insights into complicated phenomena including learning [11] and biological evolution [5] can be obtained.

Evolutionary algorithms typically generate a vast number of different agents in a short time, and it is problematic to try and analyze individual agents for a variety of reasons. Firstly, most representations are capable of encoding the same playing strategy in many ways, requiring extra effort to remove duplicates. Secondly, evolved strategies are usually highly complex, even if only from mutational drift about a central strategy and not true intricacy. Lastly, any analysis technique that works only on one representation is inherently tied to all of its biases and emphases on particular behaviors, and cannot be used for any cross-representational study. One method of solving this issue is to simplify the encoding to reduce the complexity, removing the chance of encountering interesting strategies in the process, and this is still but one choice of representation.

In [12] the notion of a *fingerprinting* operator is introduced: a representation-independent method of reducing a player's strategy into a functional form, whence standard analysis techniques can be applied [13], [14], [15]. Past applications include analysis of representational sensitivity [16], [17] and the effect of time and population size [18]. However, there are issues with using the expected score, such as possible discontinuity in fingerprints of non-finite state strategies [12], indistinguishability of very different strategies [14], and limitation to working with simultaneous two-player games.

This paper presents a general model for fingerprinting any iterated game, applicable to asymmetric or non-simultaneous games and also extensible to imperfect information or noisy games. It is developed in terms of a weighted sum over all move histories when the given player plays against a completely parametrized probabilistic finite state transducer representing the other players in the game. Free parameters include the number of states and the distribution over move lengths to consider, and can be reduced to the prior model with appropriate choices and projections.

Functional properties of the fingerprints are derived, including necessary and sufficient conditions for players to be indistinguishable, uniform convergence of approximations, analyticity and Lipschitz continuity for fingerprints of all players that are consistent with the laws of probability and equicontinuity for the entire set of bounded-state probabilistic finite state transducers. The issues in the prior model are explained here as problems with their implicit choice of length distribution, namely that the expected value is zero everywhere. Algorithms for the computation of specific cases are developed, which permit the rapid analysis of agents including identification and classification.

The theoretical results in this paper prove various properties of this general fingerprinting model; to do so rigorously, machinery from many different fields is required. Readers unfamiliar with the fundamentals of the following areas are invited to consult the references given therewith: automaton theory [19], [20]; game theory [4]; probability or measure theory [21], [22]; linear algebra with emphasis on Markov chains [23], [24]; real analysis up to function classes and multivariate Taylor series [25], [26].

The rest of the paper is organized as follows: Section II gives notation and preliminary definitions, Section III gives the definition of the fingerprint operator as well as two lemmas, and Section IV proves various properties of the operator. Example algorithms and computations for Prisoner's Dilemma, a listing of basic strategies and comparison with prior work are in Section V, while an application to discovering the structure of the space of 2-state deterministic automata, with algorithms, is presented in Section VI. Conclusions, future directions for extension, improvement and application are given in Section VII.

## II. Preliminary Definitions

*Notation:* Denote the set of natural numbers $\{0, 1, 2, \dots\}$ by $\mathbb{N}$, by $\mathbb{N}^*$ the set of positive integers $\{1, 2, 3, \dots\}$, by $\mathbb{Z}_n$ the set of natural numbers less than $n$ $\{0, 1, 2, \dots, n-1\}$, by $\mathbb{Z}_n^*$ the positive integers less than $n$ $\{1, 2, 3, \dots, n-1\}$; unless explicitly stated no reference to modular arithmetic is implied. Denote the reals by $\mathbb{R}$, the powerset operator as $\mathbb{P}$, and the characteristic or indicator function of a set $S$ by

$$\chi_S : \mathbb{U} \to \mathbb{R},\ \chi_S(x) = \begin{cases} 1 & \text{if } x \in S, \\ 0 & \text{if } x \notin S \end{cases}$$

where $\mathbb{U}$ is the universe under consideration. For an alphabet $\Sigma$, denote the set of all nonempty strings over $\Sigma$ by

$$\Sigma^+ = \bigcup_{n=1}^{\infty} \Sigma^n.$$

Denote also the set of all nonempty strings over $\Sigma$ of length at most $k$ by

$$\Sigma^{[k]} = \bigcup_{n=1}^{k} \Sigma^n.$$

*Definition 1:* Let $S$ be a nonempty set, and define $\Delta(S)$ to be the space of categorical distributions supported on subsets of $S$:

$$\Delta(S) = \big\{ v : S \to \mathbb{R} \mid v(s) \geq 0 \ \forall s \in S,\ \sum_{s \in S} v(s) = 1 \big\}$$

A $v \in \Delta(S)$ represents the probability mass function of a random variable with state space $S$. For infinite sets $S$ this may have to be a generalized function; for a finite set $S$ of cardinality $n$, this can be identified with the right $n-1$-simplex in $n$ dimensions. For $s \in S,\ v \in \Delta(S)$, denote by $v_s$ the $s$-th component of $v$, $v(s)$. As an example of this, consider $S = \mathbb{Z}_2$, which gives

$$\Delta(\mathbb{Z}_2) = \{ (x_0, x_1) \in \mathbb{R}^2 \mid x_0, x_1 \geq 0,\ x_0 + x_1 = 1 \}$$

using vector instead of functional notation. This is just the line segment from $(0, 1)$ to $(1, 0)$: for now we will, after scaling and rotation, identify it with the interval $[0, 1]$.

Recall both the definitions for a *probabilistic finite automaton* and a *finite state transducer*, and combine them.

*Definition 2:* A *probabilistic finite state transducer* (PFT) is a 5-tuple $(Q, \Sigma, \Gamma, \delta, q_0)$, where

- $Q$ is a finite nonempty set of *states*,
- $\Sigma$ is a finite nonempty *input alphabet*,
- $\Gamma$ is a finite nonempty *output alphabet*,
- $\delta : Q \times \Sigma \to \Delta(Q \times \Gamma)$ is the *transition/output function*,
- $q_0 \in \Delta(Q \times \Gamma)$ is the *initial state/output*.

Remembering that for probabilistic machines, everything is in terms of stochastic vectors, or distributions.

Note that we will mostly be working with single-state PFTs (or P1STs) whence the transition part of the function is moot. Also there is the non-standard addition of an initial output with no input seen at all: this will allow a PFT player who has to play blind to start.

*Definition 3:* An *iterated two-player perfect information game* is one where there are two abstract players $P_1$ and $P_2$, and in each round of the game, $P_1$ chooses an abstract move from the finite nonempty *moveset* $\Gamma_1$, and $P_2$ chooses from the moveset $\Gamma_2$. There is no restriction on whether the plays are simultaneous, but each player knows unambiguously their opponent's past moves. We refer to a game as the pair $(\Gamma_1, \Gamma_2)$, and define $G_1 = |\Gamma_1|, G_2 = |\Gamma_2|$.

Although here we have only two players, consider a fixed-order $n$-player game, $P_i$ having moveset $\Gamma_i$, $1 \leq i \leq n$, and the turn order is fixed. Consider the point of view of a distinguished $P_1$ (who is not necessarily the first player in the order), disregarding the special cases of the first and last move which require simple modifications. The new information available to $P_1$ on their next turn includes the current moves played by all $P_i$ before $P_1$ in the turn sequence, as well as the rest of the opponents' last moves - exactly one from each opponent. Therefore, $P_1$ while playing sees but a single "opponent" whose moveset is all $n-1$-tuples of moves of all the other players:

$$\Gamma_2^{'} = \prod_{i=2}^{n} \Gamma_i.$$

Since we are going to analyze single players who are all in the same position in the game, henceforth we only consider two player games. Lastly, we define a representation-independent characterization of a player.

*Definition 4:* Let $P$ be a player with moveset $\Gamma_1$ in the game $(\Gamma_1, \Gamma_2)$. Define the *specification* of $P$ as the function

$$\rho_P : (\Gamma_1 \times \Gamma_2)^+ \to [0, 1]$$

$$\rho_P(w,z) = \Pr(P \text{ plays } w \mid \text{opponent plays } z),$$

the probability that at each round of the game, $P$ plays the corresponding move in $w$, given that over the entire course of the game the opponent has played the move string $z$.

We consider players to be completely specified by their response probabilities to each possible history of moves. Another interpretation, supposing the player $P$ is a PFT, is that $\rho_P(w,z)$ is the probability the machine outputs $w$ given the input string $z$.

## III. Definitions and Lemmas

*Definition 5:* Given the number of states $k \in \mathbb{N}^*$ and a length distribution $\mu \in \Delta(\mathbb{N}^*)$, the *parametrized PFT probe* fingerprint operator for the game $(\Gamma_1, \Gamma_2)$ is defined as

$$\mathcal{F} : \mathcal{P} \to \big(\mathcal{M}_k \to \Delta(\Gamma_1 \times \Gamma_2)\big)$$

where $\mathcal{P}$ is the space of all players in said game that are given moveset $\Gamma_1$, and $\mathcal{M}_k$ is the space of all possible parameters for a general $k$-state PFT with input and output alphabets $\Gamma_1$ and $\Gamma_2$ respectively. Labelling the states as $Q$, that gives

$$\mathcal{M}_{|Q|} = \Delta(Q \times \Gamma_2) \times \Delta(Q \times \Gamma_2)^{|Q \times \Gamma_1|}$$

where the first term specifies the initial state/output $q_0$, then the transition/output function $\delta$. Denote by $\mathcal{F}_P$ the function output by the operator: $\mathcal{F}_P = \mathcal{F}(P)$. For $\vec{v} \in \mathcal{M}_k$ denote the PFT constructed with those parameters as $O_k(\vec{v})$; the function is defined as

$$\mathcal{F}_P(\vec{v})_{m_1 m_2} = \sum_{n=1}^{\infty} \mu(n) \sum_{\substack{(w,z)\in(\Gamma_1\times\Gamma_2)^n \\ w \text{ ends with } m_1 \\ z \text{ ends with } m_2}} \rho_{O_k(\vec{v})}(z,w)\rho_P(w,z).$$

The interpretation is that the product of $\rho$s is the two-way probability that both players in the game play as dictated by the move sequence $(w,z)$, weighting the entire set of sequences of length $n$ by $\mu(n)$, and finding the grand weighted probability that the move-pair $m_1 m_2$ occurs. To see this is well defined, the sum of all components of the fingerprint is a triple sum over entire probability distributions ($P$ must play a sequence, similarly for $O_k(\vec{v})$ and $\mu$) and hence equals 1.

For a given $\vec{v} \in \mathcal{M}_k$, denote by $r_{\emptyset q_1 m_2}$ the probability of playing $m_2$ initially and starting in state $q_1$, and $r_{k m_1 m_2}$ the probability, from state $k$ of playing $m_2$ having seen $m_1$ as input. For 1-state machines we drop the state subscript. We now have two conditions on $\mu$ which will come in as hypotheses in the theorems.

*Definition 6:* Call $\mu \in \Delta(\mathbb{N}^*)$ *nonvanishing* if

$$\forall n \in \mathbb{N}^*,\ \mu(n) \neq 0,$$

that is, it assigns positive weight to every length, and call $\mu$ *regular* if

$$\lim_{n\to\infty} \sum_{i=1}^{n} \mu(i) = 1,$$

that is the partial sums converge to 1.

For an example of a length distribution/weight function that fails these conditions, consider the (generalized) function "expectation value" which assigns equal weight to every length. By the weak law of large numbers (we are working with probabilities, *a fortiori* finite numbers) this is

$$\mu_{\mathbb{E}}(m) = \lim_{n\to\infty} \frac{1}{n} \chi_{\mathbb{Z}^*_{n+1}}(m)$$

and hence it assigns zero weight to every single length, with partial sums identically zero. This can be seen as the value of a single observation, by extension any finite number of observations, has no effect on the (infinite) direct computation of the expectation.

We will use an elementary result in multivariate power series in a forthcoming theorem:

*Lemma 1:*

$$\text{Let } f(x_1, x_2, \ldots, x_n) = \sum_{k_1,\ldots,k_n=0}^{\infty} c_{k_1,\ldots,k_n} \prod_{i=1}^{n} (x_i - \bar{x}_i)^{k_i}$$

be an $n$-variate power series centered at the point $\bar{x} = (\bar{x}_1, \ldots, \bar{x}_n)$. $f$ is identically zero in an open subset of its area of convergence that contains $\bar{x}$ if and only if $f$ is the zero power series.

*Proof:* ($\Leftarrow$) Trivial.

($\Rightarrow$) By induction. For a univariate series identically zero on an open interval, since the function is already a power series and hence trivially analytic, construct the Taylor series of $f$ centered at $\bar{x}$:

$$f(x) = \sum_{n=0}^{\infty} \frac{f^{(n)}(\bar{x})}{n!}(x - \bar{x})^n$$

where $f^{(n)}$ is the $n$th derivative of $f$. Since the function is zero on an interval we immediately see that every derivative evaluated at $\bar{x}$ is zero and hence $f$, being equal to its series, is zero.

Now consider a $k{+}1$-variate power series identically zero on an open set containing the reference point. Pick a distinguished variable $x_1$, fix everything else and construct the Taylor series of $f$ with respect to $x_1$ centered at $\bar{x}_1$:

$$f(x_1; x_2, \ldots, x_{k+1}) = \sum_{n=0}^{\infty} \frac{1}{n!} \left.\frac{\partial^n f}{\partial x_1^n}\right|_{(\bar{x}_1, x_2, \ldots, x_n)} (x_1 - \bar{x}_1)^n$$

and notice that for each $n$ (particularly $n = 0$) this is itself a power series in the $k$ variables $x_2, \ldots, x_{k+1}$. We have that $f(x_1)$ is the zero power series in $x_1$ and hence each coefficient is a zero power series in $x_2, \ldots, x_{k+1}$, and the full $f$ is the zero function on $k+1$ variables. ■

We now restrict our attention to the $k = 1$ case or that of a single-state PFT. Define first a counting function to shorten the writing a bit:

*Definition 7:* For an alphabet $\Sigma$, define the *counting function* as $C : \Sigma^+ \times \Sigma \to \mathbb{N}$,

$$C(w,a) = \text{number of occurrences of } a \text{ in } w$$

We can characterize the set of all strings indistinguishable by any P1ST as follows:

*Definition 8:* Let $(w,z) \in (\Gamma_1 \times \Gamma_2)^+$, $m_1 \in \Gamma_1$ and $m_2, m_3 \in \Gamma_2$. Define the *indistinguishable class* $[w,z]_{m_1 m_2 m_3}$

as

$$[w,z]_{m_1m_2m_3} = \left\{ \begin{array}{l} (x,y) \in \hat{S}_n(w,z) \mid x \text{ ends with } m_1, \\ z \text{ ends with } m_2 \text{ and starts with } m_3 \end{array} \right\}$$

where $\hat{S}_n(w,z)$ denotes any permutation of the (modified, see below) string pair $w,z$, considered as a single string over the alphabet $\Gamma_1 \times \Gamma_2$. That means the count of each character in $\Gamma_1 \times \Gamma_2$ must be the same.

Heavily abusing the notation, we force the pairs of moves to be ($O$'s input, $O$'s output). This means that if the game is simultaneous, then the opponent's output occurs one turn later in the game-centric perspective: for our purposes the string $w$ has to be shifted one character forward for alignment, and the two dangling moves are not permuted. A concrete list of examples set in Prisoner's dilemma, starting from the string pair $w =$ CDCD and $z =$ CCDD:

(a) $\begin{smallmatrix} & \text{C D C D} \\ \text{C C D D} & \end{smallmatrix}$ (b) $\begin{smallmatrix} & \text{D C C D} \\ \text{C C D D} & \end{smallmatrix}$ (c) $\begin{smallmatrix} & \text{C D D C} \\ \text{C C D D} & \end{smallmatrix}$

(d) $\begin{smallmatrix} & \text{D C C D} \\ \text{C D C D} & \end{smallmatrix}$ (e) $\begin{smallmatrix} & \text{C D C D} \\ \text{C D D C} & \end{smallmatrix}$ (f) $\begin{smallmatrix} & \text{C C D D} \\ \text{C C D D} & \end{smallmatrix}$

(a) is the original string pair, with the shift performed. (b) only exchanges the first two characters in $w$ and is *not* in the same indistinguishable class. (c) exchanges the last two characters in both $w$ and $z$, which is an invalid transformation that also modifies one of the dangling characters (*not* in the same class). (d) exchanges the second and third columns (the dangling C on the second row is the first column), which leaves it *in* the same class. (e) exchanges the second and fourth columns, but since the last letter of $z$ has been changed, this is *not* in the same class. (f) exchanges the third and fourth columns, and while the column containing the last character of $z$ is different, $z$'s last character is itself unchanged and this *is* in the same class.

Where $m_1, m_2, m_3$ are unambiguously specified or all triples are being considered at once, we will drop the subscripts and denote the class by the shorthand $[w,z]$. We will now see that no P1ST can separate two strings in the same indistinguishable class, in the sense given in the following lemma:

*Lemma 2:* For $k = 1$,

$$\forall [w,z],\ \forall \vec{v} \in \mathcal{M}_1,\ \forall (x_1,y_1),(x_2,y_2) \in [w,z]$$
$$\rho_{O_1(\vec{v})}(y_1,x_1) = \rho_{O_1(\vec{v})}(y_2,x_2)$$

That is, every P1ST, given an indistinguishable class, must necessarily have the same probability of outputting a string given its corresponding input, for every string pair in the class.

*Proof:* Remember that a P1ST has no state information at all (memoryless), and hence (apart from the first character) responds identically to each character. Therefore

$$\begin{aligned}\rho_{O_1(\vec{v})}(z,w) =& \Pr(O \text{ plays } z_1 \text{ initially}) \cdot \\ & \prod_{i=2}^{|w|} \Pr(O \text{ plays } z_i \text{ given input } w_i) \\ =& r_{\emptyset m_3} \prod_{m_1m_2 \in \Gamma_1 \times \Gamma_2} r_{m_1m_2}^{C((w,z),m_1m_2)}\end{aligned}$$

abusing $w_i$ and $z_i$ to count letters after required alignment, where $m_3$ is the start of $z$, collecting all like terms in the parameter space. Hence if for two strings the counts are all equal and the start characters for $z$ are the same, then $\rho_{O_1(\vec{v})}$ is identical as a function of $\vec{v}$. ∎

By even more abuse of notation, for an indistinguishable class $[w,z]$ we will denote by $\rho_{O_1(\vec{v})}[w,z]$ as the value of $\rho_{O_1(\vec{v})}$ common to all strings in the class.

## IV. Theoretical Properties

The main result of this paper is a full characterization of the classes of players indistinguishable under a 1-state opponent.

*Theorem 3:* Suppose that $k = 1$ and $\mu$ is nonvanishing. Then for two players $P_1$ and $P_2$, $\mathcal{F}_{P_1} \equiv \mathcal{F}_{P_2}$ as functions if and only if

$$\forall (w,z) \in (\Gamma_1 \times \Gamma_2)^+,\ \forall m_1 \in \Gamma_1,\ \forall m_2, m_3 \in \Gamma_2$$
$$\sum_{(x,y)\in[w,z]} \rho_{P_1}(x,y) = \sum_{(x,y)\in[w,z]} \rho_{P_2}(x,y)$$

That is, the total probability assigned to an indistinguishable class is equal for both players, for each and every class.

*Proof:* ($\Leftarrow$): Remembering that $\mathcal{F}_P$ is an absolutely convergent series for any player $P$, take the function difference

$$\mathcal{F}_{P_1}(\vec{v})_{m_1m_2} - \mathcal{F}_{P_2}(\vec{v})_{m_1m_2}$$

and split the combined sum across indistinguishable classes:

$$= \sum_{\text{all } [w,z]} \mu(|w|) \sum_{(x,y)\in[w,z]} \rho_{O_1(\vec{v})}(y,x)(\rho_{P_1} - \rho_{P_2})(x,y)$$

and by Lemma 2 we can factor out the response of the opponent:

$$= \sum_{\text{all } [w,z]} \mu(|w|) \rho_{O_1(\vec{v})}[w,z] \underbrace{\sum_{(x,y)\in[w,z]} (\rho_{P_1} - \rho_{P_2})(x,y)}_{=0 \text{ by hypothesis}}$$

and hence the functions are identical over the domain $\mathcal{M}_1$.

($\Rightarrow$): We prove its contrapositive, by showing that given that one class has a total probability difference, $\mathcal{F}_{P_1} \not\equiv \mathcal{F}_{P_2}$. Suppose for some $m_1, m_2, m_3$ as defined in Definition 8 some class has a total probability differential:

$$\exists (w,z) \in (\Gamma_1 \times \Gamma_2)^+ \;\Bigg|\; \sum_{(x,y)\in[w,z]} (\rho_{P_1} - \rho_{P_2})(x,y) \neq 0.$$

Once again split the combined sum of $(\mathcal{F}_{P_1} - \mathcal{F}_{P_2})(\vec{v})_{m_1m_2}$ across classes:

$$= \sum_{\text{all } [w,z]} \left( \mu(|w|) \sum_{(x,y)\in[w,z]} (\rho_{P_1} - \rho_{P_2})(x,y) \right) \rho_{O_1(\vec{v})}[w,z]$$

and now for each $[w,z]$ we can see again from Lemma 2 that this is a unique monomial in $\{r_{\emptyset m_j}, r_{m_i m_j}\}$. Since $\mu$ and the probability differentials for each class are constants independent of $\{r_{\emptyset m_j}, r_{m_i m_j}\}$, this is now a multivariate power series.

Invoke Lemma 1 to see that this series is identically zero if and only if every coefficient is zero. But we know that there is a class where the differential is nonzero, and that $\mu$ is nonvanishing, hence we exhibit a nonzero term in the series for a contradiction. ■

The vigilant reader will have noticed that we have not strictly fulfilled the hypotheses of Lemma 1: our implicit reference point is $\vec{0}$, which is a boundary point of the domain $\mathcal{M}_1$. To remedy this, go back to the proof of Lemma 1 and note that all that is required is for each derivative evaluated at the point be zero. Since $\vec{0}$ is a corner point in a product of right simplices, every coordinate axis through it must have one side inside $\mathcal{M}_1$. Hence every directional derivative evaluated by approaching from $\mathcal{M}_1$ must be 0. Since we know it is analytic, the derivative is equal to the limit evaluated at either side, and is still 0.

Obviously, if $\mu$ vanishes on a length $n$, then any two players can differ on their responses at move $n$ with impunity and still be indistinguishable, lessening the model's power. We now need a notion of similarity between players:

*Definition 9:* Call players $P_1$ and $P_2$ $i$-similar if

$$\forall (w, z) \in (\Gamma_1 \times \Gamma_2)^{[i]}, \ \rho_{P_1}(w, z) = \rho_{P_2}(w, z)$$

That is, their distribution-responses to the first $i$ moves are identical. We now have a result on approximating fingerprints, applicable to arbitrary $k$, number of states in the probe:

*Theorem 4:* Suppose $\mu$ is regular, and $\{P_j\}_{j=1}^{\infty}$ is a sequence of players with $P_j$ being $j$-similar to a player $P$. Then

$$\lim_{n\to\infty} \mathcal{F}_{P_n} \xrightarrow{\text{uniformly}} \mathcal{F}_P$$

*Proof:* Let $\epsilon > 0$, then since

$$\lim_{n\to\infty} \sum_{i=1}^{n} \mu(i) = 1, \ \exists N_\epsilon \in \mathbb{N}^* \ \Bigg| \ \sum_{i=1}^{N_\epsilon} \mu(i) > 1 - \epsilon$$

Now $\forall i \geq N_\epsilon$, for an arbitrary $\vec{v} \in \mathcal{M}_k$, consider the functional difference $|\mathcal{F}_{P_i}(\vec{v})_{m_1 m_2} - \mathcal{F}_P(\vec{v})_{m_1 m_2}|$

$$= \left| \underbrace{\sum_{n=1}^{\infty}}_{i\text{-similar}} \mu(n) \sum_{\substack{(w,z)\in(\Gamma_1\times\Gamma_2)^n \\ m_1 m_2 \text{ restrictions}}} \rho_{O_k(\vec{v})}(z, w)(\rho_{P_i} - \rho_P)(w, z) \right|$$

$$\leq \sum_{n=j+1}^{\infty} \mu(n) \sum_{\substack{(w,z)\in(\Gamma_1\times\Gamma_2)^n \\ m_1 m_2 \text{ restrictions}}} \underbrace{\rho_{O_k(\vec{v})}(z, w)}_{\text{sum}\leq 1} \underbrace{|(\rho_{P_i} - \rho_P)(w, z)|}_{\text{sum}\leq 1}$$

$$\leq \sum_{n=j+1}^{\infty} \mu(n) < \epsilon,$$

shrinking the absolute value by the triangle inequality. Hence $\mathcal{F}_{P_i}$ converges uniformly component-wise and therefore as a vector function to $\mathcal{F}_P$. ■

Next we give a slightly more tractable formula that computes the fingerprint of a PFT.

*Theorem 5:* Suppose $P$ is a player encodable by a PFT, and $\mu$ is regular. Then $\mathcal{F}_P$ is an analytic function over its domain.

*Proof:* We start with the construction given in [13]. Denote the machine playing $P$ by $(Q_1, \Gamma_2, \Gamma_1, \delta_1, q_{01})$ and the opponent similarly with 1s swapped with 2s. Since both players are now PFTs, we can cross-link the output of one machine to the input of the other; recall our addition of an initial output, which lets the system bootstrap itself. This is simply the construction of a product automaton which now has no input nor output: the transition probabilities from each pair of states are now constants (for a given $\vec{v}$). We also include with the pairs of states the pair of last moves, which turns this contraption into a first-order Markov chain as detailed below:

The Markov chain has as state space $Q_1 \times Q_2 \times \Gamma_1 \times \Gamma_2$ and will be indexed by state pair and move pair. The entries in the transition matrix $T$ are:

$$T_{q_1 q_2 m_1 m_2 \to q_1' q_2' m_1' m_2'} = \delta_1(q_1, m_2)_{q_1', m_1'} \delta_2(q_2, m_1)_{q_2', m_2'}$$

that is, the transition requires the events that both $P$ and $O$ decide to shift states and output correctly. Then, the fingerprint can be seen to be

$$\mathcal{F}_P(\vec{v})_{m_1 m_2} = \left\langle \chi_{m_1 m_2} \middle| \sum_{n=1}^{\infty} \mu(n) T^{n-1} \middle| q_{01} \otimes q_{02} \right\rangle$$

where the last term is the tensor product vector of the initial state/output distributions for both players, and $\chi_{m_1 m_2}$ is abusively the indicator vector that is 1 if the state-move pair indexed has move $m_1 m_2$ and zero otherwise. The entry is in the form of a real inner product.

Notice that the $n = 1$ term is simply the initial distributions, and each subsequent move has one power of the transition matrix multiplied on; the sum weighted by $\mu$ has all the components that have last move $m_1 m_2$ lifted out, as per the definition of the fingerprint operator.

Finally, it is obvious that for any finite $n$, we have that $T^n$ has polynomial entries. Since $\mu$ is regular, we use the same trick in Theorem 4, from the boundedness of entries in $T^n$, to see that the fingerprint, as a uniform limit of a series of polynomials, is itself analytic. ■

Example computations can be found further on in Section V. With this theorem, we can show that the fingerprint functions are highly well-behaved, being analytic (infinitely differentiable and equal to its power series) and hence Lipschitz continuous (since differentiable, equivalent to having a bounded derivative). These results only hold for players which are actually possible: the specification may include such absurdities as $\rho(m_1, m_2) = 0$ with $\rho(m_1 m_1, m_2 m_2) > 0$.

*Corollary 6:* Suppose $\mu$ is regular, then for a possible player $P$, $\mathcal{F}_P$ is analytic over its domain.

*Proof:* Given a player $P$ that follows the laws of probability, we can construct a PFT that is $k$-similar to it by simply listing all move sequences up to length $k$ as states in a prefix tree and constructively assigning the transition and output probabilities to match $P$; further moves are irrelevant. By Theorem 5 we see that the fingerprint of any PFT is analytic, and by Theorem 4 we have that the fingerprint of $P$ is the uniform limit of a sequence of fingerprints of PFTs, each an analytic function, and thus analytic. ■

*Corollary 7:* Suppose $\mu$ is regular, then for a possible player $P$, $\mathcal{F}_P$ is Lipschitz continuous over its domain. The precise definition is

$$\exists C \in \mathbb{R} \mid \forall x, y \in \mathcal{M}_k, \ \|\mathcal{F}_P(x) - \mathcal{F}_P(y)\| \leq C \|x - y\|.$$

*Proof:* We already know that $\mathcal{F}_P$ is analytic over the domain, so it can be expressed as a Taylor series in any basis about any point in the domain, hence $\mathcal{F}'_P$ is finite for any direction at any point. Obviously we also have $\|\mathcal{F}'_P\|$ continuous, and since the domain $\mathcal{M}_k$, being a product of simplices, is compact, $\|\mathcal{F}'_P\|$ attains a maximum on it and hence $\mathcal{F}'_P$ is bounded. Therefore

$$\|\mathcal{F}_P(x) - \mathcal{F}_P(y)\| = \left\| \int_y^x \mathcal{F}'_P(s)ds \right\| \leq C\,\|x-y\|$$

where both the integral and the derivative are in the direction of $y - x$ (remember the simplices are by definition convex) and $C$ is the bound on $\|\mathcal{F}'_P\|$, by the fundamental theorem of calculus and the mean value theorem. ■

With this, we can prove a final, and even more powerful corollary. Recall the definition of uniform continuity of a function $f : X \to Y$ as

$$\forall\epsilon > 0,\ \exists\delta > 0 \mid \forall x, y \in X, \quad \|x-y\|_X < \delta \Rightarrow \|f(x) - f(y)\|_Y < \epsilon,$$

or a single $\delta(\epsilon)$ bound works for every point in the domain. Lipschitz continuity strengthens this statement by requiring that $\delta$ be at least linear in $\epsilon$. For a set of functions, the notion of uniform equicontinuity extends this to all functions in the set at once.

*Corollary 8:* If $\mu$ is regular, then the $k$-state fingerprints of all PFTs with at most $j$ states, $\mathcal{P}_j$, form a uniformly equicontinuous set over their common domain. The precise statement is

$$\forall\epsilon > 0,\ \exists\delta > 0 \mid \forall P \in \mathcal{P}_j,\ \forall x, y \in \mathcal{M}_k, \quad \|x-y\| < \delta \Rightarrow \|\mathcal{F}_P(x) - \mathcal{F}_P(y)\| < \epsilon,$$

a single bound works for every point, for every fingerprint simultaneously.

*Proof:* We have parametrized the $k$-state opponent; since we are now considering all PFTs with at most $j$ states, we shall parametrize them as well. Let $P_j(\vec{u})$ be the parametrized player with parameter space $\mathcal{M}'_j$ similarly defined as

$$\mathcal{M}'_{|Q|} = \Delta(Q \times \Gamma_1) \times \Delta(Q \times \Gamma_1)^{|Q \times \Gamma_2|}.$$

Again construct the Markov chain for the product automaton with the same form of transition matrix, and conclude that the infinite sum-inner product form of the fingerprint holds. Repeat the rest of the proof for Theorem 5: the partial sums are bounded polynomials in all the parameters; $\mu$ regular implies that the partial sums converge uniformly to the fingerprint, which is hence analytic in every parameter.

Finally, the analogous proof from Corollary 7 shows that the set of bounded-state PT fingerprints is equi-Lipschitz, although such a characterization is uncommon. *A fortiori*, they form a uniformly equicontinuous set. ■

## V. Examples and Comparison

Prior work in this field has mostly been concerned with investigating Prisoner's Dilemma playing agents [12], [13], [14], [15]. For such a two-move game, and for $k = 1$, we can specify $O_1$ with three parameters: $x$ is the probability of cooperating initially, $y$ is the probability of cooperating in response to a cooperate, and $z$ is the probability of cooperating in response to a defect. Recall the identification of $\Delta\big(\{\mathrm{C},\mathrm{D}\}\big)$ with $[0, 1]$.

For actually computing the fingerprint, we have to specify the length distribution $\mu$: a choice that satisfies both the nonvanishment and regularity conditions is the geometric distribution:

$$\mu_\alpha(n) = (1-\alpha)\alpha^{n-1},\ \alpha \in (0,1).$$

More importantly, in [27] it is proved that computing the geometrically weighted matrix power series can be done easily with

$$\sum_{i=0}^{\infty} M^i = (I - M)^{-1},$$

the direct matrix analogue of the geometric sum. True to that, the convergence condition requires that the matrix have eigenvalues in $(-1, 1)$. Since the construction in Theorem 5 produces a Markov chain, it is column-stochastic and hence has spectral radius 1, which after scaling with $\alpha$ is strictly less than 1: the theorem applies. Hence the fingerprint of the PFT $P$ with $\mu_\alpha$ is

$$\begin{aligned}\mathcal{F}_P &= \left\langle \chi_{m_1 m_2} \left| \sum_{n=1}^{\infty} (1-\alpha)(\alpha T)^{n-1} \right| q_{01} \otimes q_{02} \right\rangle \\ &= (1-\alpha)\chi^T_{m_1 m_2}(I - \alpha T)^{-1}(q_{01} \otimes q_{02})\end{aligned}$$

with $T$ the transition matrix as defined in Theorem 5.

This is computable in the time to invert a matrix multinomial, which is at most $\mathcal{O}(d^m n^{3+m})$ where $m$ is the number of variables and $d$ is the maximum degree of entries, by inverting the matrix at sufficient points in an $m$-dimensional grid and interpolating. Each inversion takes $\mathcal{O}(n^3)$, and since the entries in the inverse are $(2n-1)$ bounded-degree rational expressions of the entries in the input matrix (which are themselves polynomials in the variables), $\mathcal{O}(dn)$ points are required per dimension.

For our purposes $m = kG_2(kG_1 + 1)$ and $d = 3$ (product of $\alpha$, the initial move and output distribution for the opponent, all linear), hence $6n - 2$ points to invert at per dimension are sufficient by brute force.

### A. Display Details

We display fingerprints of Prisoner's Dilemma strategies as they are conceptually easier to understand, as well as in line with previous work [13], [14], [15]. We will use the space $[0, 1]^3$ to represent $\mathcal{M}_1$.

Since $x$, the probability of initially cooperating, affects things once, we can see that the fingerprint function is linear in it; hence two points suffice to completely determine it: we will conveniently choose $x = 1$ (cooperate initially) and $x = 0$ (defect initially).

In forthcoming figures, we will depict the $yz$-plane from $y = 0$ (bottom) to $y = 1$ (top), and $z = 0$ (left) to $z = 1$ (right). Recall also that the output of the fingerprint function is a probability vector over the pairs of moves: there are 4 dimensions, but one is redundant since they sum to 1, which conveniently fits in the RGB color model.

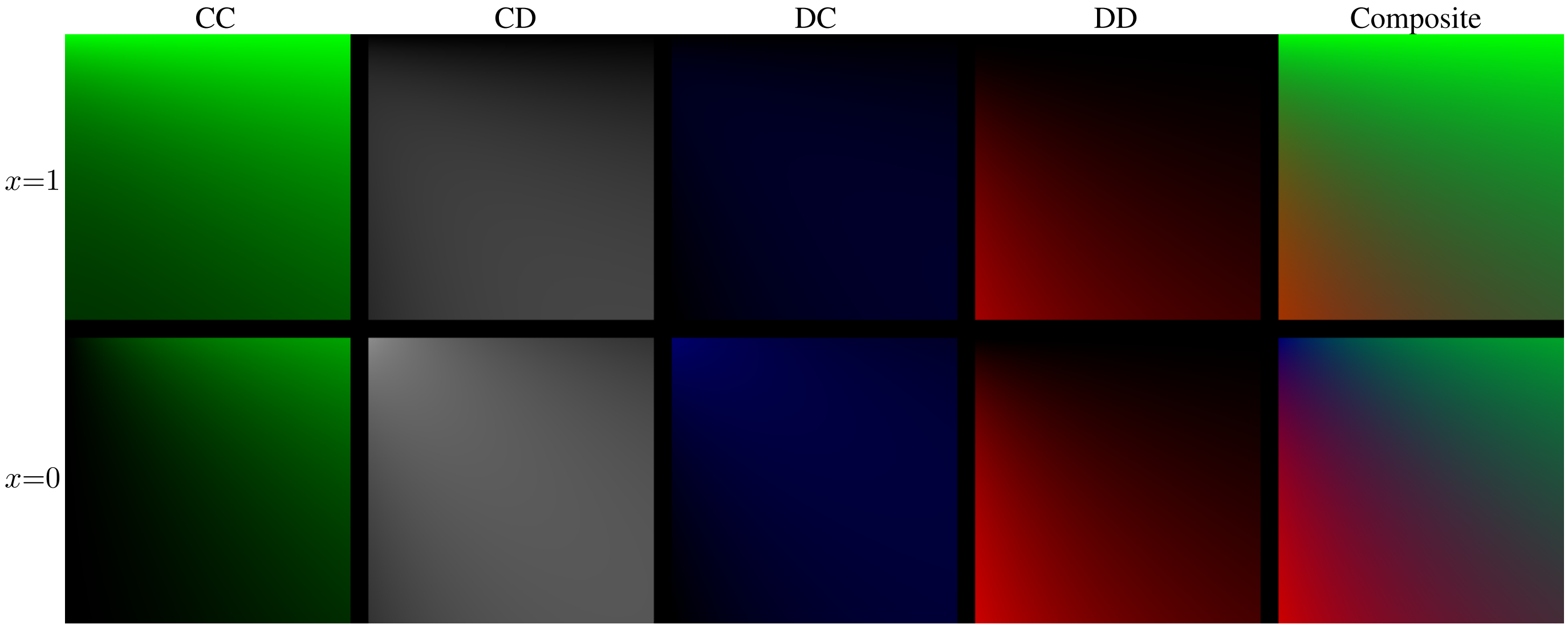

Fig. 1. Fingerprint panels for **TFT** "Tit-for-tat" (cooperate initially, return opponent's last move) at $\alpha = 0.8$, see Subsection V-A for more detail

The figures displayed have 10 panels; the upper row is for $x = 1$ and the lower row $x = 0$. In each row, the first panel (green) shows the probability of CC (both cooperate), linearly from $0 \mapsto (0, 0, 0)$ RGB to $1 \mapsto (0, 1, 0)$. The second panel (white) shows the probability of CD (player $P$ cooperates, opponent probe defects), linearly from $0 \mapsto (0, 0, 0)$ to $1 \mapsto (1, 1, 1)$. The third panel (blue) shows Pr(DC) (player $P$ defects, opponent probe cooperates) linearly from $0 \mapsto (0, 0, 0)$ to $1 \mapsto (0, 0, 1)$; the fourth panel (red) shows Pr(DD) (both defect) linearly from $0 \mapsto (0, 0, 0)$ to $1 \mapsto (1, 0, 0)$. The last panel is a composite that adds the colors from the green, blue and red panels, $(\text{Pr(DD)}, \text{Pr(CC)}, \text{Pr(DC)})$ RGB; the white panel is ignored, so the darker the color, the more CD occurs.

The geometric $\mu$ defined in the previous section will be the one of choice; the value $\alpha = 0.8$ is chosen for most of the figures arbitrarily. The "mean" of the distribution is $\frac{1}{1-\alpha}$, and not to give undue weight to the initial move, $\alpha$ close to 1 is preferred.

### B. Tit-for-Tat Computation

As an example, we will compute the fingerprint of the famous strategy tit-for-tat (cooperate initially, return the opponent's last move thereafter). Following the construction in Theorem 5, we impute a Markov chain to the product automaton as follows:

$$\begin{array}{ccccc} & 11CC & 11CD & 11DC & 11DD \\ 11CC & y & 0 & z & 0 \\ 11CD & 1-y & 0 & 1-z & 0 \\ 11DC & 0 & y & 0 & z \\ 11DD & 0 & 1-y & 0 & 1-z \end{array}$$

Where the Markov states are labelled (TFT's state, probe's state, TFT's last move, probe's last move). Call this matrix $T$.

The fingerprint with geometric $\mu$ is given by:

$$\mathcal{F}_{\text{TFT};m_1m_2} = \chi^T_{m_1m_2}(I_4 - \alpha T)^{-1} \begin{bmatrix} x \\ 1-x \\ 0 \\ 0 \end{bmatrix},$$

where the indicator vectors are given by $\chi_{CC} = e_1$, $\chi_{CD} = e_2$, $\chi_{DC} = e_3$, $\chi_{DD} = e_4$, the standard basis vectors. Intermediate calculations are omitted; the four component functions are given in the next column.

The astute reader may notice that the choice of $e_i$, the natural basis vectors, for the initial move vector corresponds to fixing the initial move from both automata; computing the fingerprint of a variant of TFT that defects first move is as simple as replacing that vector with $(0, 0, x, 1 - x)^T$. In this sense we have computed the fingerprint of all possible choices of initial moves (and states) from both players.

$$\mathcal{F}_{\text{TFT;CC}} = \frac{x\big(1-\alpha(1-z)\big) + \alpha^2(z-x)\big((1-\alpha)y + \alpha z\big)}{\big(1+\alpha(z-y)\big)\big(1+\alpha^2(z-y)\big)}$$

$$\mathcal{F}_{\text{TFT;CD}} = \frac{\big(1-\alpha(1-z)\big)\big((1-x)+\alpha(x-y)\big)}{\big(1+\alpha(z-y)\big)\big(1+\alpha^2(z-y)\big)}$$

$$\mathcal{F}_{\text{TFT;DC}} = \frac{\alpha\big((1-\alpha)y+\alpha z\big)\big((1-x)+\alpha(x-y)\big)}{\big(1+\alpha(z-y)\big)\big(1+\alpha^2(z-y)\big)}$$

$$\mathcal{F}_{\text{TFT;DD}} = \frac{\alpha\big((1-y)(1-x)+\alpha(x-y)(1-y)\big)}{\big(1+\alpha(z-y)\big)\big(1+\alpha^2(z-y)\big)}$$

The reader should also verify that $\lim_{\alpha\to 0^+}$ removes dependence on $y, z$ as we have $\mu = \chi_{\{1\}}$ and hence responses do not matter; $\lim_{\alpha\to 1^-}$ removes dependence on $x$ as we are now specifying $\mu = \mu_{\mathbb{E}}$, the expected value, and hence initial move does not matter. Also verify that this second limit, with the parameter substitution derived further on in Section V-E, reduces the function to the one given in [13]. This fingerprint function, evaluated at $\alpha = 0.8$, is displayed in Figure 1.

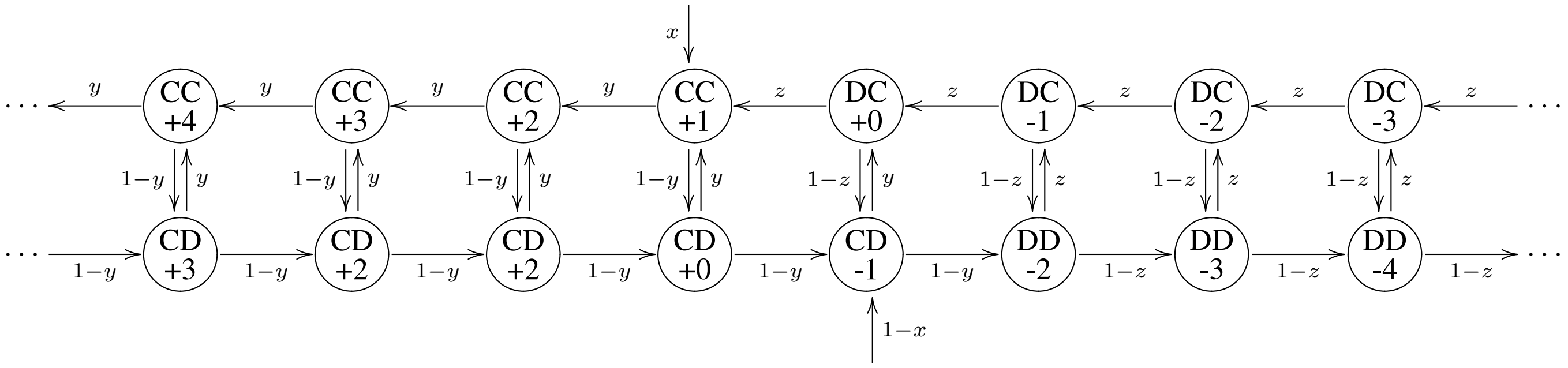

Fig. 2. Schematic representation of the Markov chain construction for computing the fingerprint of Majority. States are labelled Majority's last move, probe's last move as well as current cooperations minus defects.

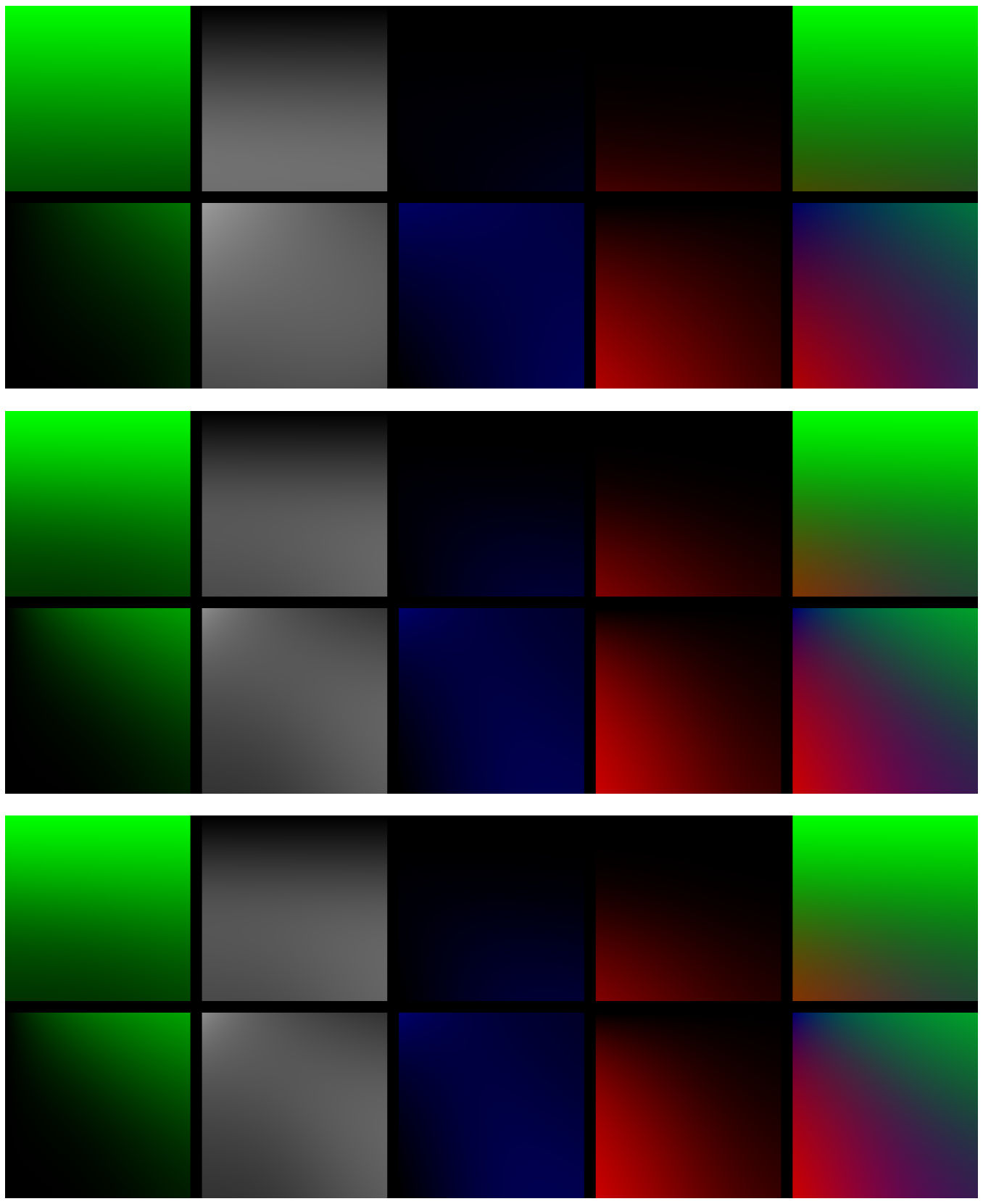

Fig. 3. Approximate fingerprints for Majority truncated at 4 (top), 16 (middle) and 256 (bottom) moves, with the geometric distribution at $\alpha = 0.8$

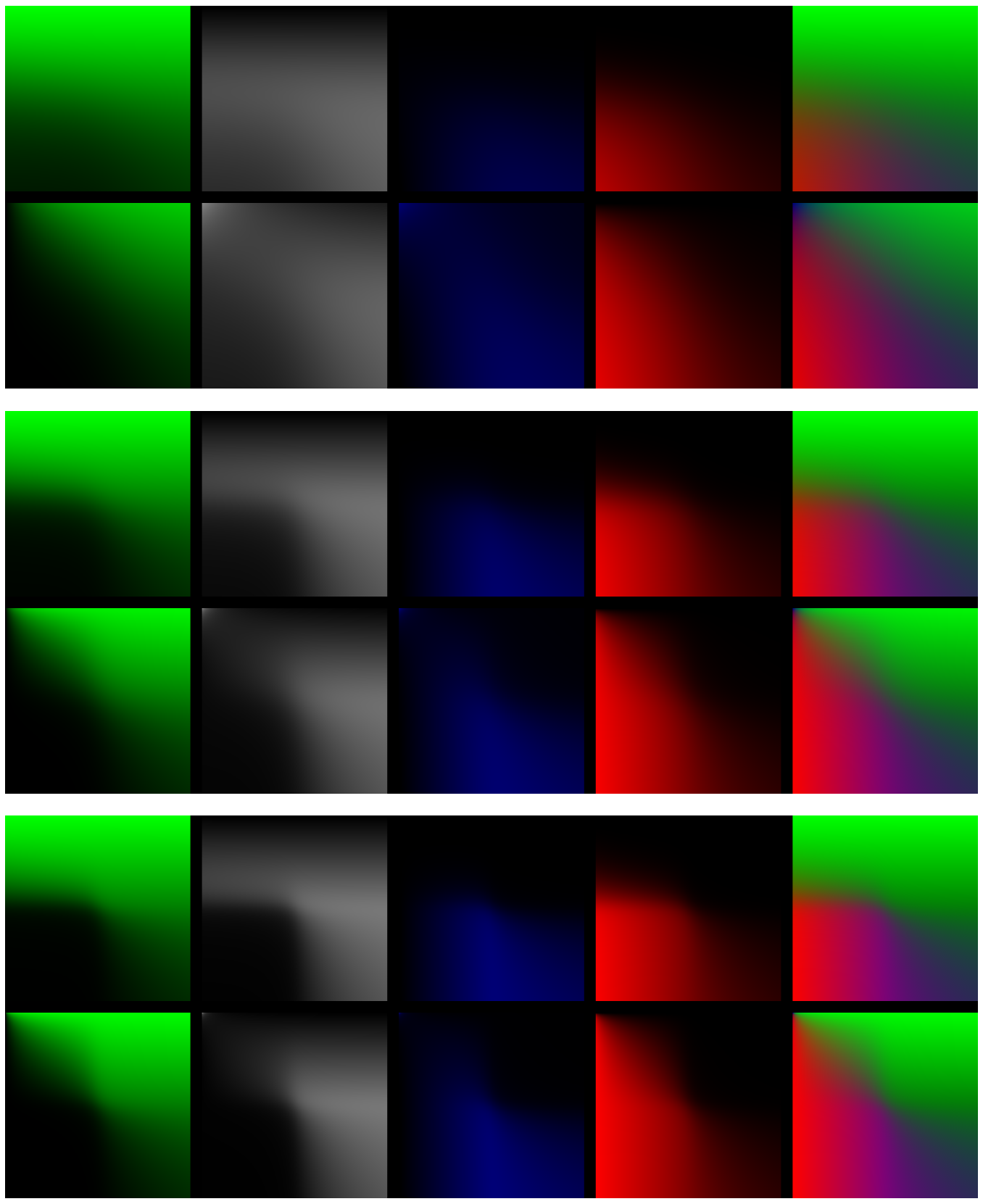

Fig. 4. Approximate fingerprints for 8192-move truncated Majority with the geometric distribution at $\alpha = 0.9$ (top), 0.98 (middle) and 0.995 (bottom)

### *C. Majority Computation*

Majority is the strategy that returns the move the opponent used more, breaking ties in favor of cooperation. This requires unbounded memory and hence this is not implementable as a finite state machine. While the result of Corollary 6 shows that it is analytic for reasonable $\mu$, actually computing it is another matter. The result of Theorem 5 gives a theoretically feasible way: construct a sequence of automata that play identically to majority for the first $n$ moves, compute their fingerprints and take the limit. We will use an equivalent approach (minus the limit).

Consider the product automaton used for computing fingerprints between an infinite state machine that plays majority and the one-state probe automaton. A schematic representation of the Markov chain extension is shown in Figure 2. The states are labelled with a pair of last moves and a number, which counts cooperation minus defects from the opponent; the probe's state is never in question.

We may truncate this Markov chain with a finite bound on the number allowed as states. If the bound is $\pm n$, then it takes at least $n$ moves to reach the end. Truncating the infinite sum definition of the fingerprint at the same bound, we get:

$$\mathcal{F}_{\text{Majority}} \approx (1-\alpha)\chi^T_{m_1 m_2}\left(\sum_{i=1}^{n}(\alpha T)^{n-1}\right)(q_{01}\otimes q_{02}).$$

This is conceptually a "strategy" that plays as majority up to the $n$th move then flatly refuses to play - yet this is still $n$-similar to majority.

To compute this, we may track the probability mass as it runs through the truncated Markov chain and hence directly compute $T^k q_i$ for $k = 0, \ldots, n-1$. Results for $\alpha = 0.8$ for $n = 4, 16, 256$ are shown in Figure 3. A bound on the absolute error in terms of the probability mass omitted is $\alpha^{n+1}$ and hence the sequence converges linearly in the number of states. The bound is on the order of $10^{-25}$ at $n = 256$, negligible.

To illustrate the result in [12] that majority's expected value fingerprint is discontinuous, we vary $\alpha$ from $0.9, 0.98, 0.995$ in Figure 4 to $0.999$ in Figure 5. To keep the error bound reasonably low, we increase $n$ to 8192, giving the largest error at $0.999^{8193} \approx 10^{-4}$. Notice the sharpening corner at the point $y = z = 0.5$.

The observant reader may have noticed that unlike TFT, whose fingerprint function becomes independent of $x$ as $\mu$ approaches the expected value, here the $x = 0$ and $x = 1$ tracks seem to converge towards something different. This is a consequence of the proof in [12] that shows the fingerprint "type" depends on the relative probability that the Markov chain diverges towards $+\infty$ or $-\infty$. In the region where $y \in (0.5, 1], z \in [0, 0.5)$ (opponent plays close to TFT), the "equilibrium point" about 0 is unstable and the initial move tips the balance irrevocably towards one mode of divergence over another. This shows another defect in using the expected value: while one may expect that for an automaton with no transient states initial moves do not matter and can be ignored, in some infinite cases it may still have permanent effects!

### D. *Catalogue of Simple Strategies*

A selection of simple strategies is fingerprinted and displayed all with geometric $\mu$ at $\alpha = 0.8$. In the top half: "Always cooperate" (ALLC, Figure 6a) never defects and hence its DC/DD panels are empty, likewise with "Always defect" (ALLD, Figure 6b) and its CC/CD panels. Compare "Tit-for-tat" (TFT, Figure 1) and "Psycho" (Figure 6c): their composite panels are approximately reverses. "Random" (Figure 6d) has differences in color since the opponent does not always play randomly. "Periodic CD" (Figure 6e) and "Periodic DC" (Figure 6f) are distinct, owing to the geometric distribution emphasizing earlier moves: Periodic CD is closer to the ALLC fingerprint.

In the bottom half: "Pavlov" (Figure 6g) and "Fortress-2" (Figure 6h) play exact reverses of each other, but interestingly in expected value their fingerprints match - here they are distinguishable. Fortress-2 also has its cooperative side revealed by forcing an initial defect from the probe. This is limited however, and "Fortress-3" (Figure 6j) shows a fingerprint close to ALLD. These two are examples of a handshake or password strategy - to activate their cooperation, you have to play the password (a defect or two defects). "Vengeful" (Figure 6i) plays like Always defect if the opponent defects first; while continued mutual cooperation can persist. "Tit-for-two-tats" (Figure 6l) is a more forgiving variant of TFT, while "Two-tits-for-tat" (Figure 6k) is more vengeful, and it shows.

### E. *Comparison*

A comparison with work by Ashlock and Kim [14]: their fingerprinting model focuses on first specifying the finite state strategy and then adding random noise to it, while this model starts at a probabilistic finite machine and goes on from there. Their fingerprint using the strategy TFT as a probe can be computed as follows:

$$F_{\text{TFT}}(P, x, y) = \mathbb{E}(\text{score of } P \text{ against } JA(\text{TFT}, x, y))$$

the expected score of playing against a parametrized opponent. $JA(\text{TFT}, x, y)$ is, if $x + y \leq 1$: plays cooperate with probability $x$, defect with probability $y$ and as TFT would otherwise; if $x + y \geq 1$: plays cooperate with probability $1 - y$, defect with probability $1 - x$ and the reverse of what TFT would otherwise.

This is a lengthy definition, but we can try and reparametrize it in terms of P1ST parameters $(x', y', z')$. If $x + y \leq 1$, the probability of initially playing cooperate is if the machine picks either randomly or to follow TFT: $x' = x + (1 - x - y) = 1 - y$; the probability of cooperating in response to a cooperate is either random or as TFT: $y' = 1 - y$; and cooperation in response to a defect can only occur randomly: $z' = x$. If $x + y \geq 1$, the initial cooperation can only be random: $x' = 1 - y$; cooperation in response to a cooperate is also random-only: $y' = 1 - y$; cooperation in response to a defect is either random or as reverse-TFT: $z' = 1 - y + \big(1 - (1 - y) - (1 - x)\big) = x$.

Hence in the entire unit square, we have that

$$JA(\text{TFT}, x, y) = O_1(1 - y, 1 - y, x)$$

and we get immediately the proof at [13] that this fingerprint is a continuous extension of the same function in the $x + y \leq 1$ triangle to the entire unit square. We see that using TFT as a probe is a two-dimensional subspace of the full P1ST parameter space, omitting an independent variable for the initial move; similarly using a $k$-state strategy gives a 2-dimensional subspace of $\mathcal{M}_k$. Also instead of outputting a probability vector of move-pairs, they use it to compute a score, reducing it to a real-valued function.

Since their model is subsumed into this one, with $\mu$ being the expectation value, this new model is a strict generalization of theirs. Additionally, recall that expectation fails both nonvanishment and regularity, conditions for some of the theorems. Recall [12] where the strategy Majority is shown to have a discontinuous fingerprint, illustrated in Figure 5.

## VI. Metrics and Structure of DFT-Space

One application of the fingerprinting method is to investigate how distinct strategies differ from one another, and by extension how different representations sample the space of players. Recall from Definition 1 that the set of distributions supported on a $n$-element set is isomorphic to the surface of the $n-1$-simplex. This gives us a natural way of equipping it with a measure:

*Definition 10:* Let $S$ be a finite nonempty set with $|S| = n$. Let $(\Delta(S), \Sigma_S, \pi_n)$ be the probability space defined with:

$$\pi_n(X) = \frac{(n-1)!}{\sqrt{n}} \lambda_{n-1}\big(P(X)\big)$$

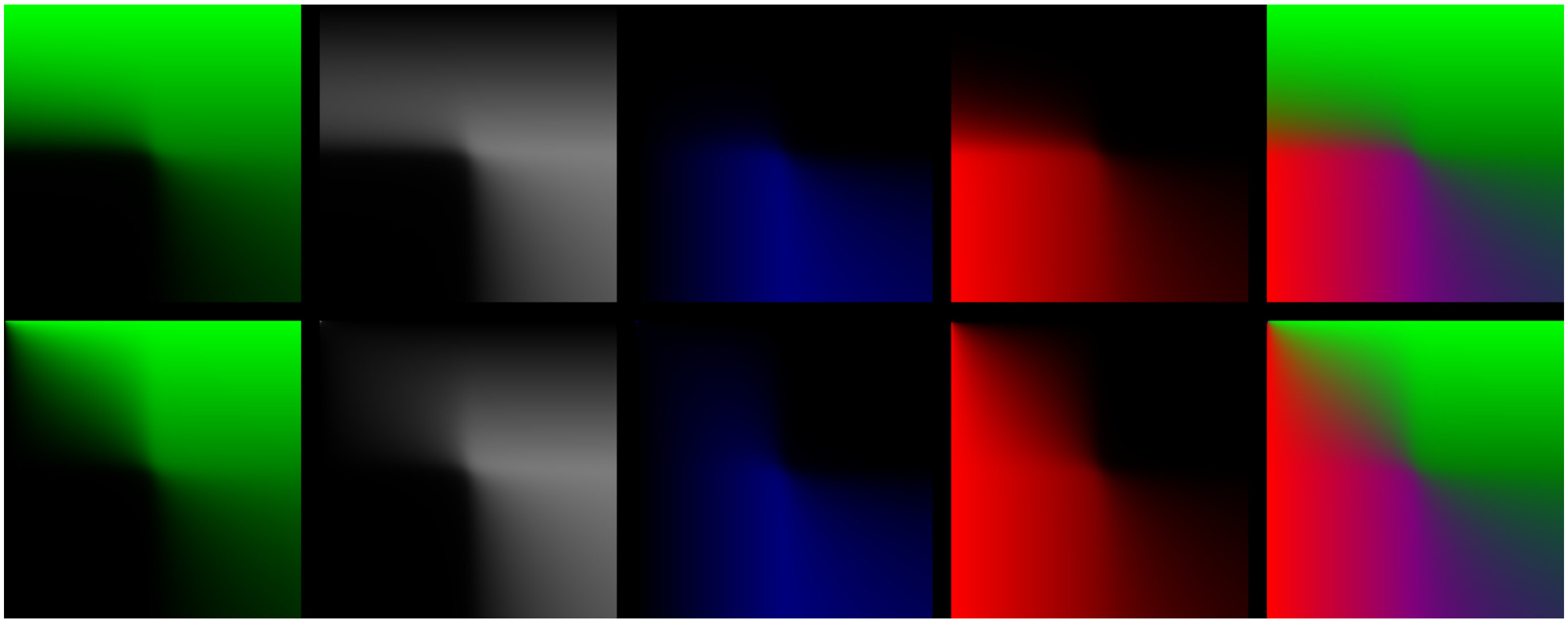

Fig. 5. Approximate fingerprint for 8192-move truncated Majority with the geometric distribution at $\alpha = 0.999$

Where $P : \Delta(S) \to \mathbb{R}^{n-1}$ is an appropriate projection that is a Euclidean isometry on $\Delta(S)$ considered as a subset of $(\mathbb{R}^n, d_2)$, and $\lambda_{n-1}$ is the standard Lebesgue measure in $n-1$ dimensions. The measurable subsets $\Sigma_S$ are just the preimages under $P$ of the Lebesgue measurable sets in $\mathbb{R}^{n-1}$, $\Sigma_{\mathbb{R}^{n-1}}$, that lie completely in the range $P(\Delta(S))$:

$$\Sigma_S = P^{-1}\Big(\Sigma_{\mathbb{R}^{n-1}} \cap \mathbb{P}\big(P(\Delta(S))\big)\Big).$$

This convoluted definition is just a mathematically precise way of stating that the measure of some subset of the surface of the $n-1$-simplex is found by projecting it onto a base, which is a solid $n-2$-simplex (for example, the unit regular tetrahedron is projected onto the solid unit right triangle), and finding its $n-1$-hyperarea.

Note that unlike a usual probability space, this one has elements in its state space that are themselves distributions. The normalizing factor is the reciprocal of the $n-1$-hyperarea of the $n-1$-right simplex. Recall that we identified $\Delta(\mathbb{Z}_2)$ with $[0, 1]$, the latter obviously having measure (length) 1. The reader can easily verify that

$$\big(\Delta(\mathbb{Z}_2), \Sigma_{\mathbb{Z}_2}, \pi_2\big) \cong \big([0, 1], \Sigma_{\mathbb{R}} \cap \mathbb{P}([0, 1]), \lambda_1\big)$$

as measure spaces. With a measure in hand, we can reproduce the (normalized) $\mathcal{L}_2$ or Euclidean distance between two functions:

*Definition 11:* Let $k$ and $\mu$ be given. Define a metric $d_2 : \mathcal{P} \times \mathcal{P} \to \mathbb{R}$, where

$$d_2(P_1, P_2) = \sqrt{\int_{\mathcal{M}_k} \|\mathcal{F}_{P_1} - \mathcal{F}_{P_2}\|_2^2 \, d\mu_{\mathcal{M}_k}} \text{ , with } d\mu_{\mathcal{M}_k} = d^{kG_1+1}\pi_{kG_2}.$$

That this is a metric (and even a norm; all fingerprints trivially have norm 1) follows immediately from $\mathcal{L}_2$ being a metric (and norm), since we have only scaled it by a constant. By inspection we can see that the maximum distance is $\sqrt{2}$ since $\|\mathcal{F}_{P_1} - \mathcal{F}_{P_2}\|_2^2$ is bounded above by 2. The attainable maximum is less than that since the opponent will not always respond with a definite move.

## A. Separation of Variables

Although the fingerprints have complex dependencies on the parameters for responses, they are linear in the initial parameters. Hence this part of the fingerprint can be separated out to reduce the dimension a bit. Rewrite the fingerprint for player $P$ as the linear combination

$$\mathcal{F}_P(\vec{s}, \vec{r}) = \sum_{q \in Q} \sum_{m_3 \in \Gamma_2} r_{\emptyset q m_3} f_{P;q,m_3}(\vec{r})$$

where $\vec{s}, \vec{r}$ are the initial state/move and response parameters respectively. Now rewrite the definition of $\mathcal{L}_2^2$ as

$$\int_{\Delta(Q \times \Gamma_2)} \int_{\mathcal{R}_k} \left\| \sum_{q \in Q} \sum_{m_3 \in \Gamma_2} r_{\emptyset q m_3} (f_{P_1;q,m_3} - f_{P_2;q,m_3})(\vec{r}) \right\|^2 d\pi_{kG_2} \, d\mu_{\mathcal{R}_k}$$

where $\mathcal{R}_k$ is the parameter space for responses and $\mu_{\mathcal{R}_k}$ its associated measure. Split the squared norm into a sum over components:

$$= \sum_{\substack{m_1 \in \Gamma_1 \\ m_2 \in \Gamma_2}} \int_{\Delta(Q \times \Gamma_2)} \int_{\mathcal{R}_k} \left( \sum_{q \in Q} \sum_{m_3 \in \Gamma_2} (f_{P_1} - f_{P_2})_{q,m_3}(\vec{r})_{m_1 m_2} \, r_{\emptyset q m_3} \right)^2 d\pi_{kG_2} \, d\mu_{\mathcal{R}_k},$$

next double the inner sums to get rid of the square:

$$= \sum_{\substack{m_1 \in \Gamma_1 \\ m_2 \in \Gamma_2}} \int_{\Delta(Q \times \Gamma_2)} \int_{\mathcal{R}_k} \left( \sum_{\substack{q_1 \in Q \\ q_2 \in Q}} \sum_{\substack{m_3 \in \Gamma_2 \\ m_4 \in \Gamma_2}} (f_{P_1} - f_{P_2})_{q_1,m_3}(\vec{r})_{m_1 m_2} \, (f_{P_1} - f_{P_2})_{q_2,m_4}(\vec{r})_{m_1 m_2} \, r_{\emptyset q_1 m_3} r_{\emptyset q_2 m_4} \right) d\pi_{kG_2} \, d\mu_{\mathcal{R}_k},$$

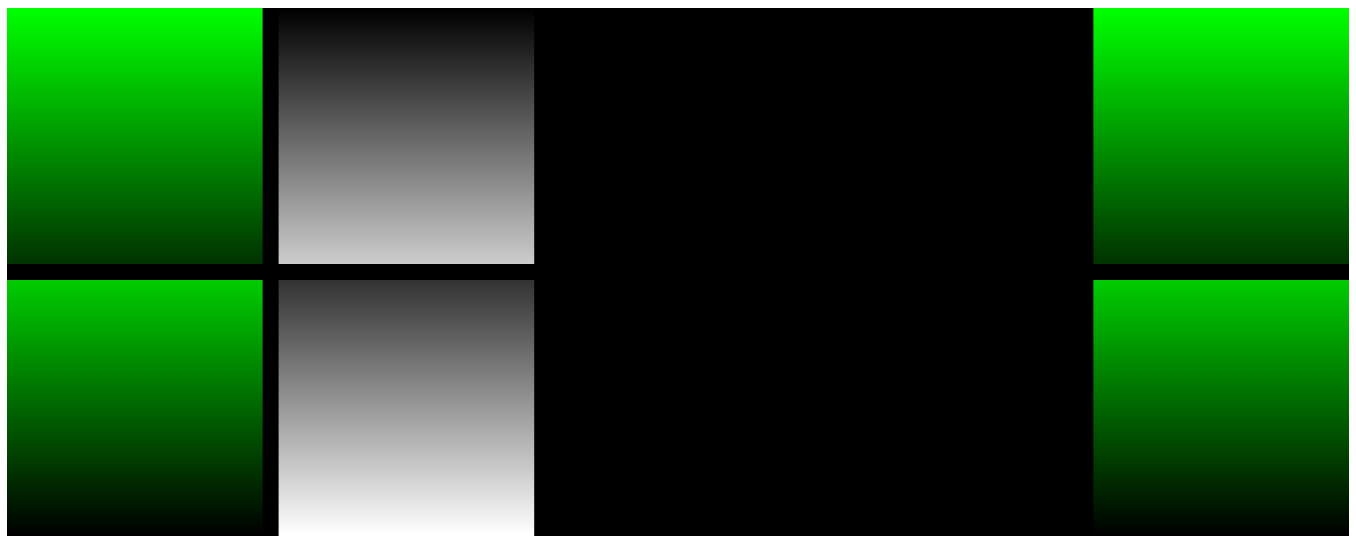

(a) **ALLC** “Always cooperate” (always cooperate)

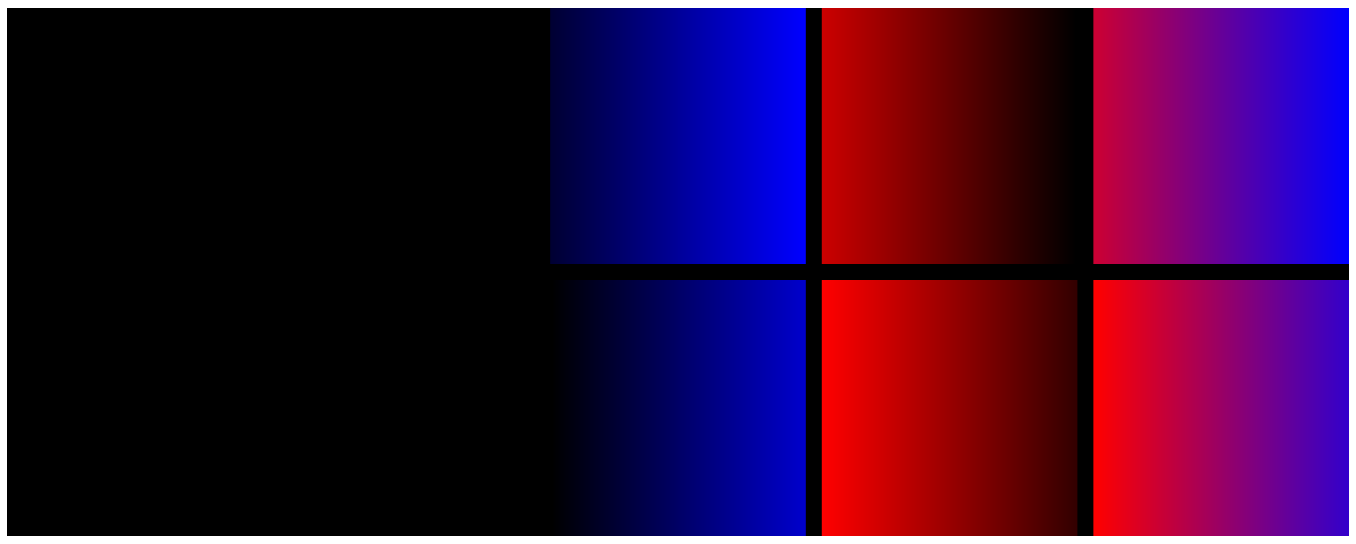

(b) **ALLD** “Always defect” (always defect)

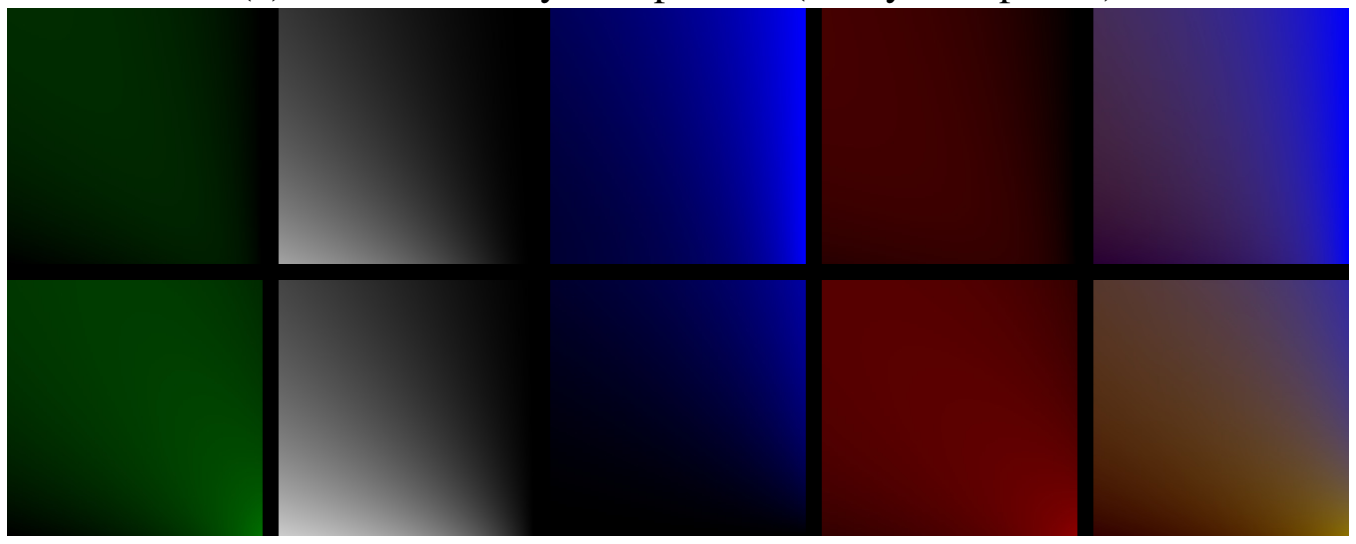

(c) **PSY** “Psycho” (defect initially, return the reverse of opponent’s action)

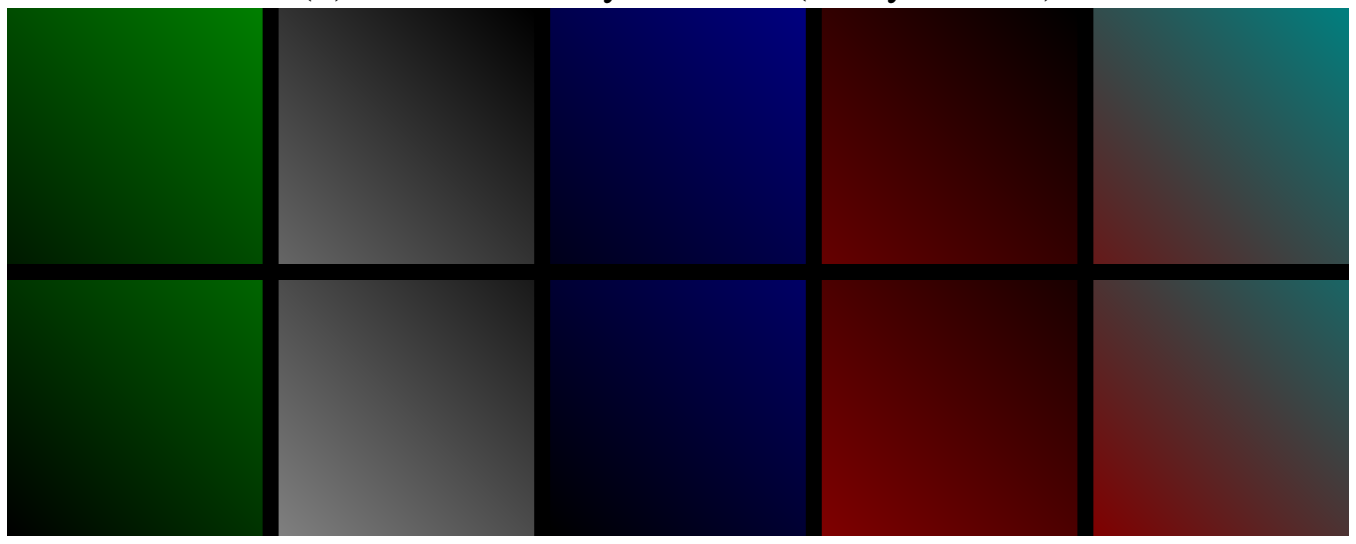

(d) **RAND** “Random” (cooperate with probability 1/2 on every move)

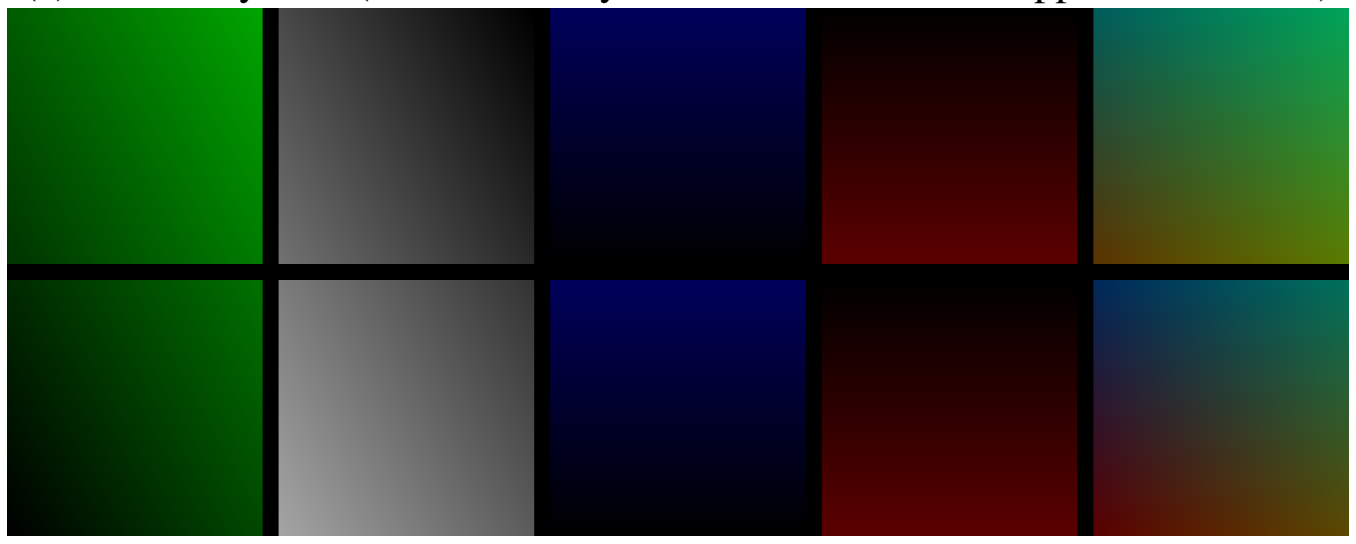

(e) **perCD** “Periodic CD” (repeat the sequence of moves C,D)

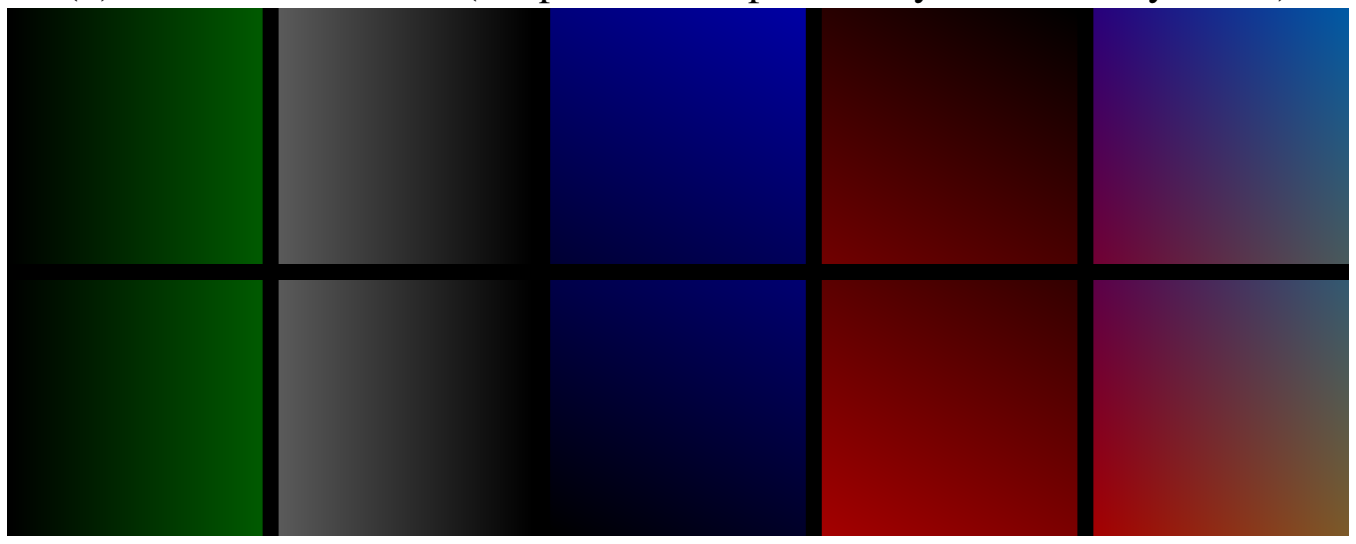

(f) **perDC** “Periodic DC” (repeat the sequence of moves D,C)

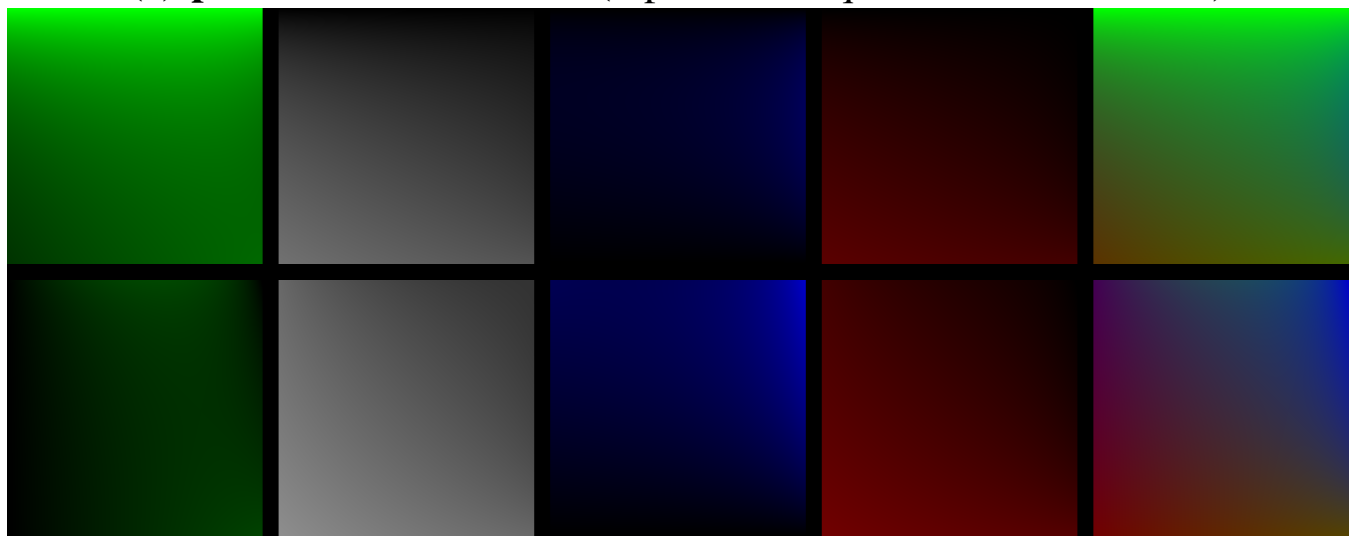

(g) **PAV** “Pavlov” (cooperate initially, then if both players’ last moves match)

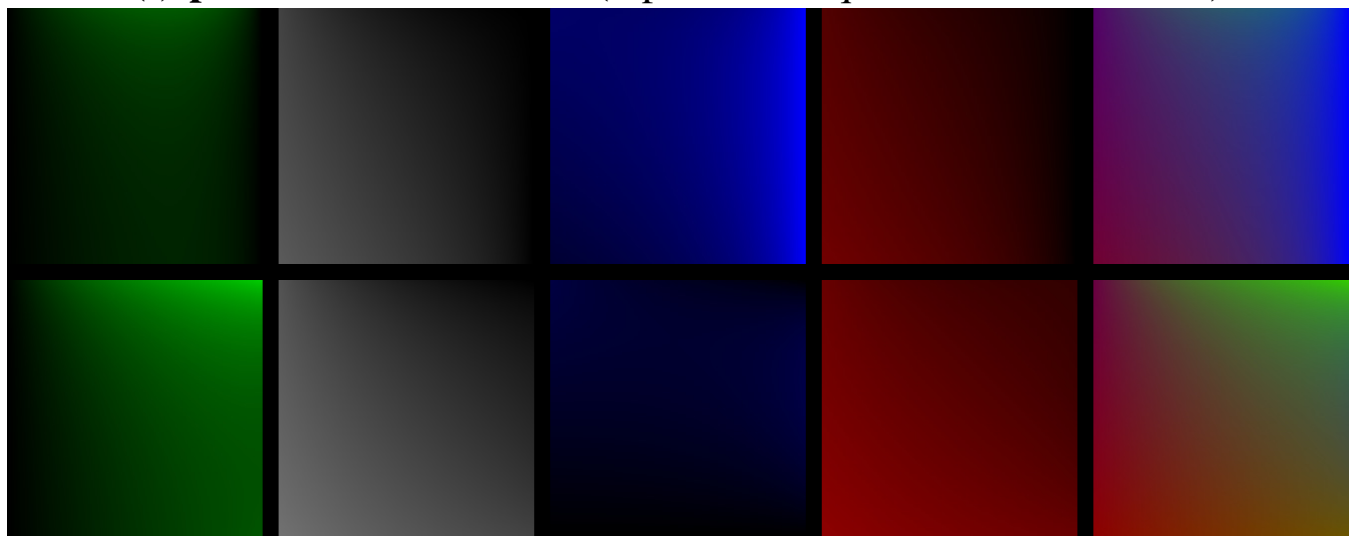

(h) **FRT2** “Fortress-2” (defect until opponent defects, then continue cooperating with cooperation, else reset)

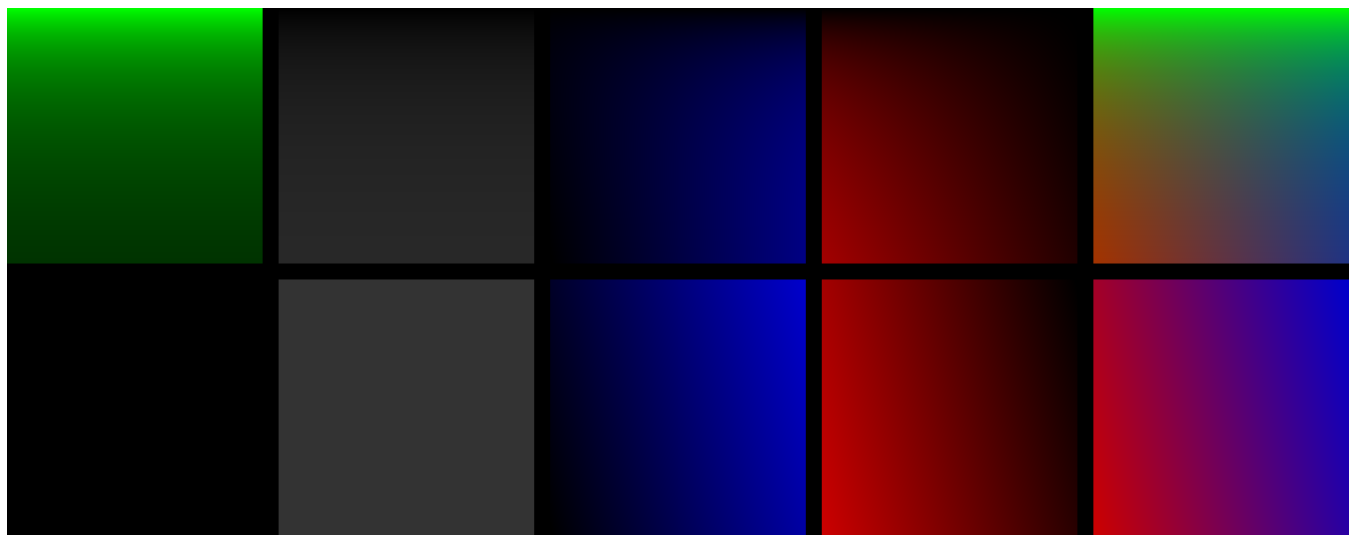

(i) **VNGFL** “Vengeful” (cooperate until opponent’s first defection, defect thereafter)

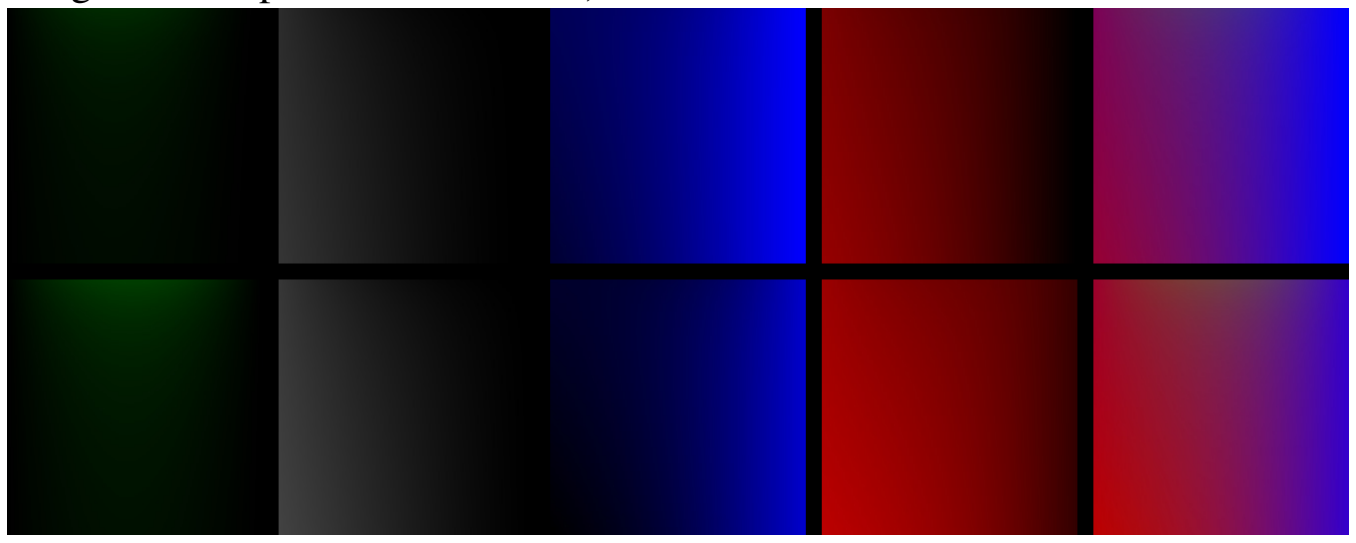

(j) **FRT3** “Fortress-3” (defect until opponent defects twice in a row, then continue cooperating with cooperation, else reset)

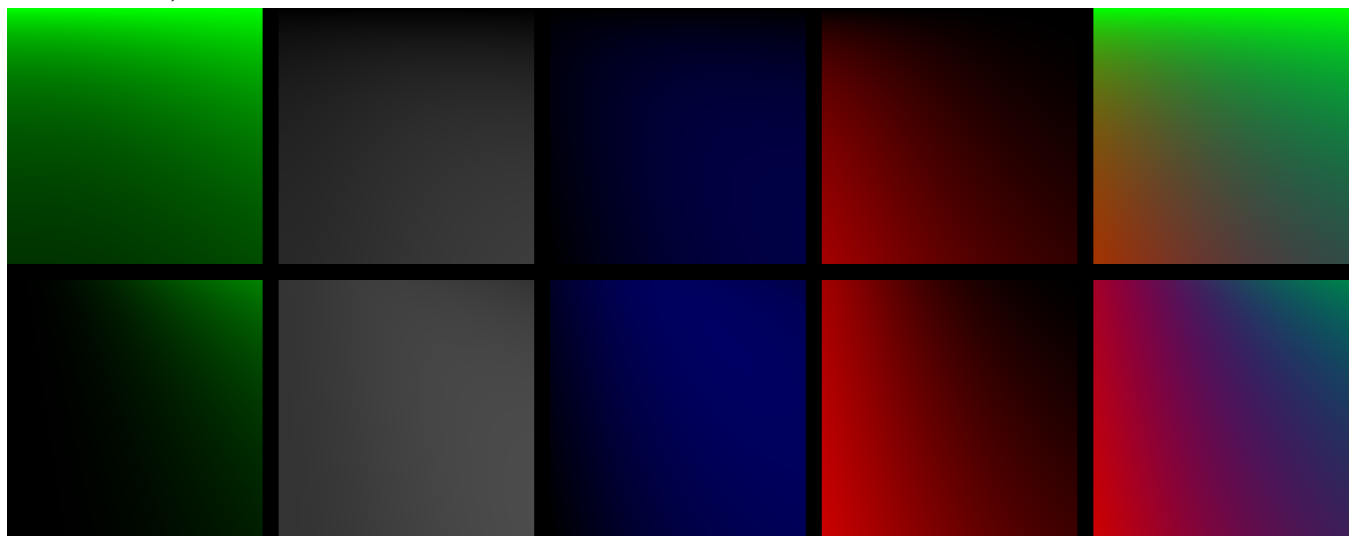

(k) **2TFT** “Two-tits-for-tat” (defect if opponent has defected in either of the last two moves)

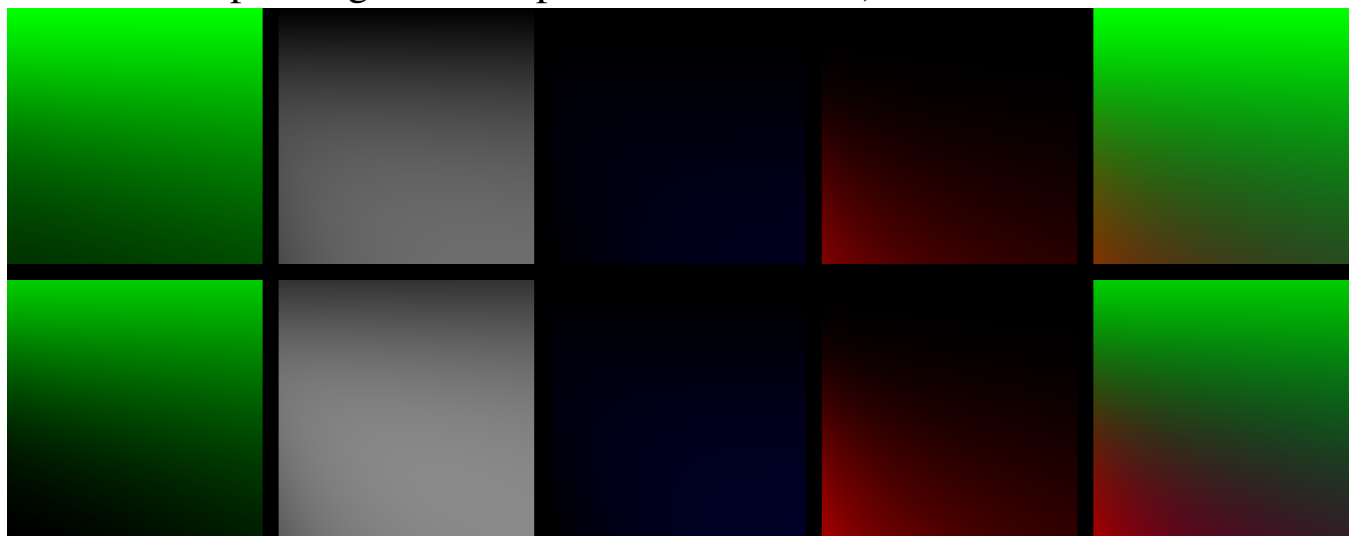

(l) **TF2T** “Tit-for-two-tats” (defect if opponent has defected in both of the last two moves)

Fig. 6. A selection of simple strategies, fingerprinted with geometric $\mu$ at $\alpha = 0.8$. The abbreviated names for each strategy are given in boldface.

which allows us to complete the separation of variables:

$$= \sum_{m_1 m_2} \sum_{q_1 q_2} \sum_{m_3 m_4} \left( \int_{\Delta(Q\times\Gamma_2)} r_{\emptyset q_1 m_3} r_{\emptyset q_2 m_4} \, d\pi_{kG_2} \right) \left( \int_{\mathcal{R}_k} \big((f_{P_1} - f_{P_2})_{q_1,m_3} (f_{P_1} - f_{P_2})_{q_2,m_4}\big)(r)_{m_1 m_2} \, d\mu_{\mathcal{R}_k} \right).$$

The quadratic integral can be easily calculated. Since the simplices are symmetric under any exchange of variables, there are only two cases depending on if $(q_1, m_3) = (q_2, m_4)$. If so, call that variable $x$ and the number of variables $kG_2 = n$, and project the surface of the right $n$-simplex into the solid right $n-1$-simplex, which scales it by $1/\sqrt{n}$. Write the integral as

$$(n-1)! \int_0^1 x^2 \int_{\blacktriangle(Q\times\Gamma_2\backslash\{x\};1-x)} 1 \, d\sigma_{n-2} \, dx$$

where $d\sigma_{n-2}$ is the unscaled measure on $n-2$ dimensions and $\blacktriangle(Q \times \Gamma_2\backslash\{x\}; 1-x)$ is the solid $n-2$-right simplex with variables in $Q \times \Gamma_2\backslash\{x\}$ of side length $1-x$, whose hypervolume scales as the $n-2$th power of length:

$$= (n-1)! \int_0^1 x^2 \frac{1}{(n-2)!} (1-x)^{n-2} \, dx,$$

and substitute $y = 1-x$:

$$= (n-1) \int_0^1 (1-y)^2 y^{n-2} \, dy = \frac{2}{n(n+1)}.$$

If not, call the two variables $x, y$, project the surface similarly and write the integral as

$$(n-1)! \int_0^1 x \int_0^{1-x} y \int_{\blacktriangle(Q\times\Gamma_2\backslash\{x,y\};1-x-y)} 1 \, d\sigma_{n-3} \, dy \, dx$$

and similarly the $n-3$ dimensional hypervolume scales as the $n-3$th power:

$$= (n-1)! \int_0^1 x \int_0^{1-x} y \frac{1}{(n-3)!} (1-x-y)^{n-3} dy \, dx,$$

and perform the change of variables $(x, y) \mapsto (x, z = 1-x-y)$ of Jacobian -1:

$$= (n-1)(n-2) \int_0^1 x \int_0^{1-x} (1-x-z) z^{n-3} \, dz \, dx$$

$$= (n-1)(n-2) \int_0^1 x \frac{(1-x)^{n-1}}{(n-2)(n-1)} \, dx,$$

and substitute $y = 1-x$:

$$= \int_0^1 (1-y) y^{n-1} dy = \frac{1}{n(n+1)},$$

which can be made the same as the equal case by merging the terms for $x, y$ and $y, x$.

Hence we have the reduced formula for the distance computation as

$$d_2(P_1, P_2) = \frac{2}{kG_2(kG_2+1)} \sum_{m_1 m_2} \sum_{(q_1,m_3)\leq(q_2,m_4)} \left( \int_{\mathcal{R}_k} \big((f_{P_1} - f_{P_2})_{q_1,m_3} (f_{P_1} - f_{P_2})_{q_2,m_4}\big)(r)_{m_1 m_2} \, d\mu_{\mathcal{R}_k} \right)$$

for some arbitrary ordering of $Q \times \Gamma_2$; $m_1 \in \Gamma_1$, $m_2 \in \Gamma_2$, $q_1, q_2 \in Q$, $m_3, m_4 \in \Gamma_2$.

## B. Approximation

Even if $k = 1$ and $\mu$ is geometric, exactly computing the distances between fingerprints is very difficult: to illustrate, the distance between TFT and ALLC as computed by the Maple mathematical package, and hand simplified beyond the capability of both Maple and MATLAB, is

$$\begin{aligned} d_2(\mathrm{TFT}, \mathrm{ALLC}; \mu_\alpha)^2 = \frac{1}{45\alpha^6(\alpha-1)^3} \big(\alpha^4(\alpha-1)(15\alpha^6 \\ - 45\alpha^5 + 226\alpha^4 - 425\alpha^3 + 487\alpha^2 - 288\alpha + 102) - (\alpha^{10} \\ - 20\alpha^9 - 20\alpha^8 - 100\alpha^7 + 110\alpha^6 - 82\alpha^5 - 110\alpha^4 + 340\alpha^3 \\ - 515\alpha^2 + 330\alpha - 78)\ln(\alpha+1) - (7\alpha^{10} + 40\alpha^9 - 110\alpha^8 \\ + 390\alpha^7 - 690\alpha^6 + 1080\alpha^5 - 1190\alpha^4 + 1060\alpha^3 - 695\alpha^2 \\ + 330\alpha - 78)\ln(\alpha^2+1) - (\alpha-1)^5(45\alpha^4 - 15\alpha^5 + 15\alpha^3 \\ + 35\alpha^2 + 60\alpha + 78)\ln(1-\alpha)\big) \end{aligned}$$

with $\ln$ denoting the natural logarithm. Both time and complexity constraints preclude analytical computation for all but the simplest strategies, hence an approximation to the distance is needed. A simple grid over the unit square will suffice, however an estimate of the error is needed.

Recall Corollary 8 which proves equicontinuity for the set of $j$-state automaton fingerprints. What we need here is a bound on the common Lipschitz constant for the set. Again use the inner product definition of the fingerprint with geometric $\mu$, define $A = (I - \alpha T)$ and derive with respect to $y$ (also $z$):

$$\frac{\partial \mathcal{F}_{P;m_1 m_2}}{\partial y} = (1-\alpha)\chi^T_{m_1 m_2} \frac{\partial}{\partial y}\left(A^{-1}\right)(q_{01} \otimes q_{02})$$

since $q_{01}$ does not depend on $y$. Using the formula $\frac{dA^{-1}(t)}{dt} = -A^{-1}\frac{dA(t)}{dt}A^{-1}$ (proved by $A^{-1} = 2A^{-1} - A^{-1}AA^{-1}$) and taking the norm we get

$$\left|\frac{\partial \mathcal{F}_{P;m_1 m_2}}{\partial y}\right| = \left|(1-\alpha)\chi^T_{m_1 m_2} A^{-1} \frac{\partial}{\partial y}(A) A^{-1}(q_{01} \otimes q_{02})\right|$$

which with linearity of the matrix norm with respect to the vector norm, submultiplicativity and the Cauchy-Schwarz inequality becomes

$$\leq (1-\alpha) \left\|\chi_{m_1 m_2}\right\|_1 \left\|A^{-1}\right\|_1^2 \left\|\frac{\partial}{\partial y} A\right\|_1 \left\|q_{01} \otimes q_{02}\right\|_1.$$

Since the last vector is a probability vector its 1-norm is 1. The $\chi$ vector has exactly $j$ entries of 1 and hence has 1-norm $j$. Recall from the definition of $T$ to see that the partial derivatives have absolute sum 2 (the entire column may contain at most $y$ and $1-y$ in combination), hence the partials of $A$ have 1-norm $2\alpha$. In [27] it is proved that the 1-norm of $A^{-1}$ is $\frac{1}{1-\alpha}$. Putting this together gives

$$= (1-\alpha)(j)\frac{1}{(1-\alpha)^2}(2\alpha)(1) = \frac{2j\alpha}{1-\alpha}.$$

Call this bound $L_y$ and notice that it diverges as $j \to \infty$ or $\alpha \to 1^-$. An identical bound works for the partial with respect to $z$: $L_z = \frac{2j\alpha}{1-\alpha}$.

For the partials with respect to $x$, only $q_{01}$ depends on it, so the norm-derivative becomes

$$\left|\frac{\partial \mathcal{F}_{P;m_1 m_2}}{\partial x}\right| = \left|(1-\alpha)\chi^T_{m_1 m_2} A^{-1} \frac{\partial}{\partial x}(q_{01} \otimes q_{02})\right|,$$

again after manipulation of the norms and inner products

$$\leq (1-\alpha)\left\|\chi_{m_1m_2}\right\|_1\left\|A^{-1}\right\|_1\left\|\frac{\partial}{\partial x}(q_{01}\otimes q_{02})\right\|_1 .$$

Again the partial of $q$ with respect to x has absolute sum 2, giving

$$=(1-\alpha)(j)\frac{1}{1-\alpha}(2)=2j.$$

Call this bound $L_z$. We will however not need this bound as we have separated out the $x$ variable and computed it exactly.

Finally, take the full distance function and express it as:

$$d_2(P_1,P_2)^2=\frac{1}{3}\sum_{m_3\leq m_4}\int_0^1\int_0^1\sum_{m_1m_2}(\Delta f_{m_3})(\Delta f_{m_4})\,dy\,dz$$

with $\Delta f_m$ being shorthand for $(f_{P_1;m}-f_{P_2;m})(r)_{m_1m_2}$. Analyze one double integral as follows. Let the functional differences $(\Delta f_m)$ be approximated by $(\Delta f_m+\delta_m)$. Expand the product to get

$$\sum_{m_1m_2}(\Delta f_{m_3}+\delta_{m_3})(\Delta f_{m_4}+\delta_{m_4})-(\Delta f_{m_3})(\Delta f_{m_4})=$$
$$\sum_{m_1m_2}\delta_{m_3}(\Delta f_{m_4})+(\Delta f_{m_3})\delta_{m_4}+\delta_{m_3}\delta_{m_4}$$

and restrictions can come in. $|\Delta f_m|\leq 1$ since they are differences of probability vectors; $\sum_{m_1m_2}\delta_m=0$ since we approximate with points on the function. There are 3 $m_3m_4$ components and we can bound those additively. Hence we have a bound on the error of the integrand

$$\epsilon\leq 12\delta_{\max}+6\delta_{\max}^2$$

and the error in the entire integral is bounded by

$$|\Delta d_2(P_1,P_2)|\leq\sqrt{\int_0^1\int_0^1 4\delta_{\max}(y,z)+2\delta_{\max}(y,z)^2\,dy\,dz}\ .$$

Suppose we divide the unit square into an $n\times n$ evenly spaced grid of squares and sample at midpoints. From the Lipschitz constant computed earlier, the bound is $\delta_{\max}(x,y,z)=\|(L_x\delta x,L_y\delta y,L_z\delta z)\|_2$ where $(\delta x,\delta y,\delta z)$ is the vector distance to the nearest grid point. Since $\delta x=0$ as we do not need to sample $x$, and $L_y=L_z$ we now have $\delta_{\max}(y,z)=L_y\|(\delta y,\delta z)\|_2$. Plug into the integral for a single square:

$$\int_{x_i}^{x_{i+1}}\int_{y_l}^{y_{l+1}}4\delta_{\max}(y,z)+2\delta_{\max}(y,z)^2\,dy\,dz$$
$$=4\int_0^{\frac{1}{2n}}\int_0^{\frac{1}{2n}}4L_y\|(y,z)\|_2+2L_y^2\|(y,z)\|_2^2\,dy\,dz.$$

The squared term is easy to compute:

$$=4\int_0^{\frac{1}{2n}}\int_0^{\frac{1}{2n}}2L_y^2(y^2+z^2)\,dy\,dz=8L_y^2\int_0^{\frac{1}{2n}}\frac{z^2}{2n}+\frac{1}{24n^3}\,dz$$
$$=\frac{L_y^2}{3n^4}.$$

The linear term naturally is moved to polar coordinates:

$$=4(4L_y)\left(\int_0^{\frac{\sqrt{2}}{2n}}\int_0^{\frac{\pi}{2}}rr\,d\theta\,dr-2\int_{\frac{1}{2n}}^{\frac{\sqrt{2}}{2n}}\int_0^{\sec^{-1}(\frac{r}{2n})}rr\,d\theta\,dr\right)$$

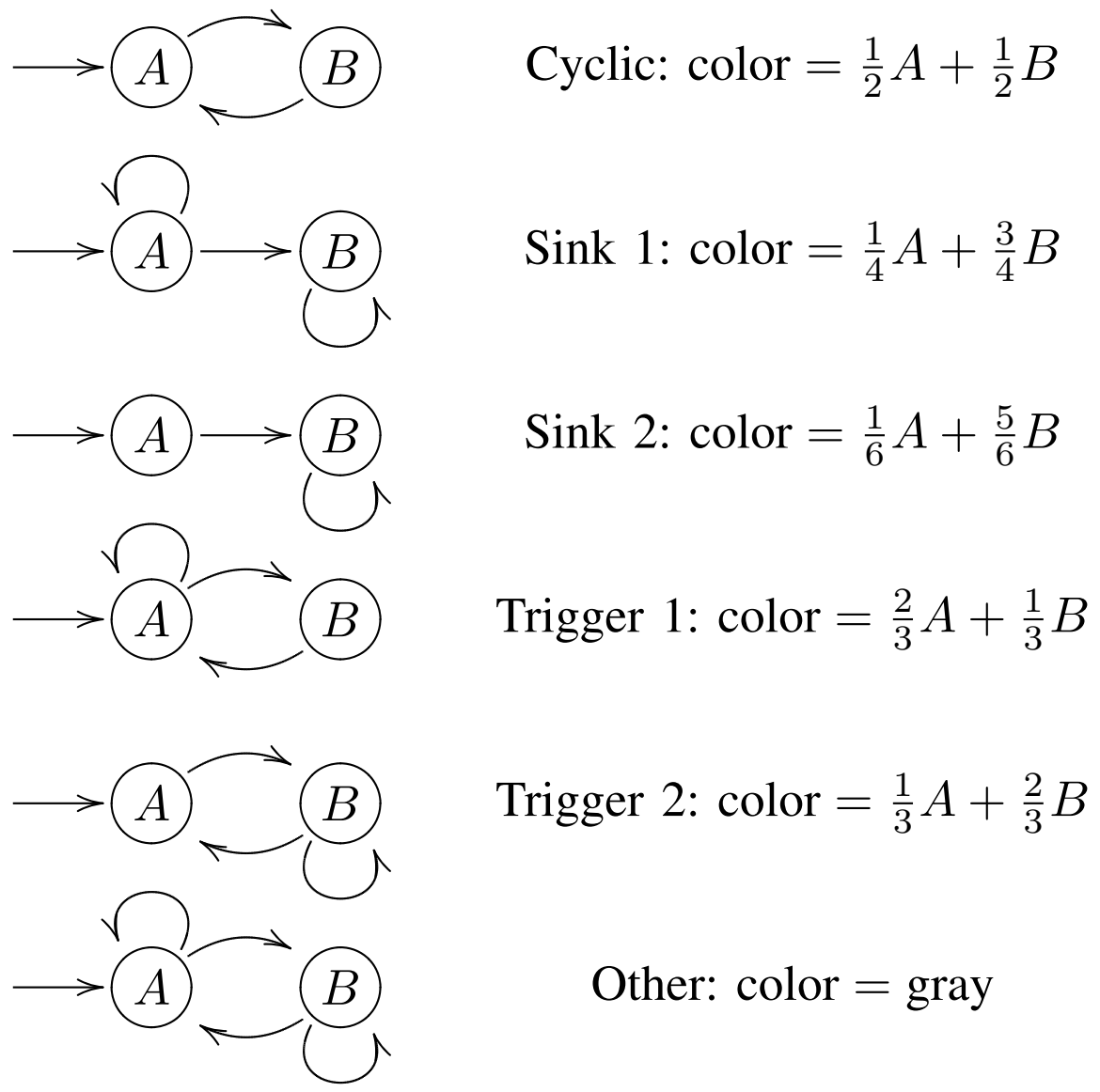

Fig. 7. Types of 2-state automata classified by transition graphs alone. $A$ and $B$ refer to strategies as played by that state disregarding transitions, and their assigned colors.

where $\sec^{-1}$ is the arcsecant function, by overintegrating the quarter circle to radius $\sqrt{2}\times$ and subtracting off the excess. The first portion of this is simple:

$$=\int_0^{\frac{\sqrt{2}}{2n}}\frac{\pi}{2}r^2\,dr=\frac{\pi}{2}\frac{1}{3}\left(\frac{\sqrt{2}}{2n}\right)^3=\frac{\pi}{12\sqrt{2}n^3}$$

After scaling $r$ by $2n$, the extraneous part becomes:

$$=2\int_1^{\sqrt{2}}\frac{1}{2n}\left(\frac{r}{2n}\right)^2\sec^{-1}(r)\,dr=\frac{1}{4n^3}\int_1^{\sqrt{2}}r^2\sec^{-1}(r)\,dr$$

which can be seen after some calculus to be

$$=\frac{1}{4n^3}\frac{1}{3}\big(\sqrt{2}(\pi-1)-\ln(1+\sqrt{2})\big).$$

The entire linear term is now

$$=16L_y\frac{1}{12\sqrt{2}n^3}\left(\pi-\sqrt{2}(\sqrt{2}(\pi-1)-\ln(1+\sqrt{2}))\right)$$
$$=\frac{2\sqrt{2}\big(2-\pi+\sqrt{2}\ln(1+\sqrt{2})\big)L_y}{3n^3}\ .$$

Lastly, since there are $n^2$ squares, the entire error bound is

$$|\Delta d_2(P_1,P_2)|\leq$$
$$\sqrt{\frac{4j^2\alpha^2}{3(1-\alpha)^2n^2}+\frac{4\sqrt{2}j\alpha\big(2-\pi+\sqrt{2}\ln(1+\sqrt{2})\big)}{3(1-\alpha)n}}\ .$$

This error is linear in $j$, diverges as $\mathcal{O}(\frac{\alpha}{1-\alpha})$ and only $\mathcal{O}(n^{-\frac{1}{2}})$. The extra work to tighten the bound is because of this divergence in $\alpha$ which cannot be easily removed.

### *C. Experimental Design*

As the space of deterministic finite state automata with given input and output alphabets and bounded states forms one of the basic finite sets of "computing machinery", we will investigate its structure as a simple example of actually applying this model in practice. The idea is to transform each member of this set into a function and do some analysis on the resultants.

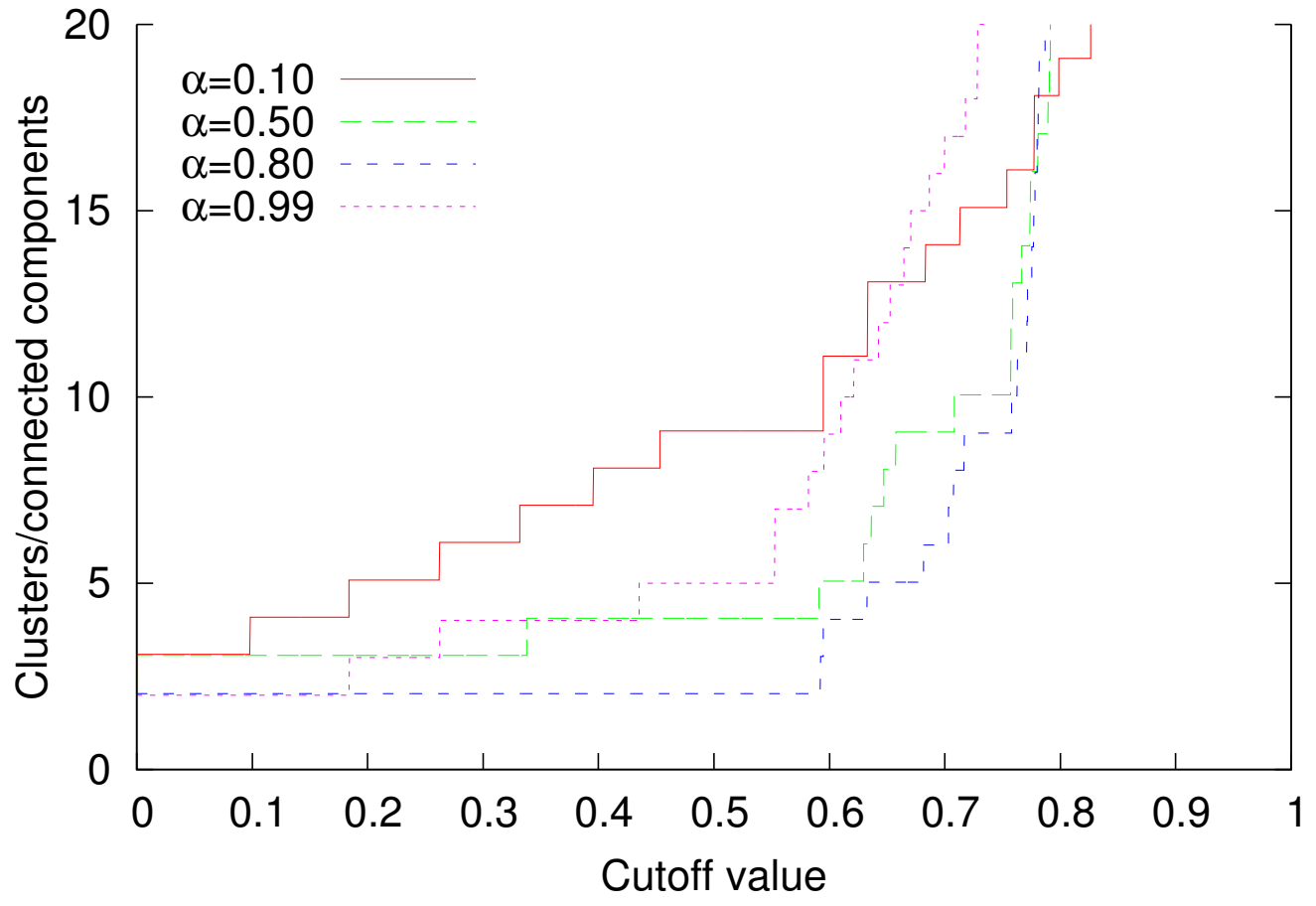

Fig. 8. Number of clusters found for data computed at 4 different $\alpha$ values against cutoff value.

We will consider the fingerprints of all 2-state DFTs for playing Prisoner's Dilemma. Nominally, there are $2^9$ of them, the binary choices being initial move, and transition and output for both inputs for both states. However, if the automaton is to have a 2-state minimal representation then restrictions appear, such as the second state has to be accessible, and the two states play different strategies from the 1-state list {ALLC,ALLD,TFT,PSY}. There are 296 automata under complete playing equivalence, 8 of which are the 1-state automata.

Each automaton is fingerprinted with the standard 1-state probe, and all pairwise distances computed approximately with a grid of $4096 \times 4096$ points over the unit square, sampled at midpoints and calculated in a binary divide-and-conquer fashion to reduce floating-point error. $\alpha$ ranges in intervals of 0.05 from 0 to 0.5, then in intervals of 0.02 from 0.5 to 0.9, then in intervals of 0.005 from 0.9 to 0.995, and at 0.99999999, a total of 51 points.

Standard statistical analysis techniques are employed to interpret the data, including $k$-means multiclustering [12] and principal components analysis [28].

Recall that standard $k$-means clustering is stochastic, depending on the choices of initial clusters. For such little data one single run of $k$-means gives unacceptably large uncertainty. Hence $k$-means multiclustering is used. This runs $k$-means many times for some distribution of cluster number and outputs a $[0, 1]$ similarity matrix based on the frequency data points $i, j$ are found in the same cluster. Reinterpreting that as a weighted graph, we may construct a clustering algorithm by removing edges below a cutoff value and calling connected components clusters. Raising the cutoff and removing more edges can only split current clusters; hence it is hierarchical.

### *D. Results and Discussion*

The error bounds computed in Section VI-B give values of 0.0098 at $\alpha = 0.5$, 0.0299 at $\alpha = 0.9$, 0.1126 at $\alpha = 0.99$ and 56382 at $\alpha = 1 - 10^{-8}$. The last bound is worthless and all data from that frame may well be meaningless, apart from observed continuity.

Figure 8 shows the results of the $k$-means multiclustering. The data for $\alpha = 0.1, 0.5, 0.8, 0.99$ are clustered 1000000 times each with uniformly random initial cluster centers, the number of such picked from a distribution which is 2 plus a geometric distribution of $p = 0.1$ truncated at 148, yielding a range of $[2, 150]$ clusters.

Notice that the 0.1 data has evenly spaced steps from 3 to 9 then a long flat region, indicating 9 important clusters with unequal spacing. Above that larger jumps occur indicating equal spacing amongst parallel subclusters. The 0.5 data has 4 important clusters, two of which are much closer than the other two, and splits into a cloud of equally spaced subclusters around cutoff 0.6 and has no further structure. The 0.8 data has two clusters, which decomposes into 9 lesser clusters, which in turn have little substructure. The 0.99 data has 2-5 major clusters and no other significant level of structure.

Principal components analysis was run using MATLAB software for each of the $\alpha$ values individually, as well as for all 51 frames concatenated. This introduces significant bias towards higher $\alpha$ in representation, but subsampling evenly makes little change to the results.

In the following figures all 296 automata are plotted at once: to give some clarity a coloring scheme was introduced. The special 1-state automata are plotted as follows: ALLC in green, ALLD in red, TFT in blue, PSY in black. Some of the 2-state automata can be considered to "interpolate" between 1-state strategies and are given colors to match. Cyclic automata are given a color halfway between the two strategies played. Sink automata are given a color 3/4 or 5/6 of the sink depending on if it is possible to stay in the transient state. The last class of automata recognizable are those that have one state always transition to the other (for our purposes called "trigger" automata): these are given 1/3 of the color of the lesser state, and further desaturated halfway to gray. The rest of the automata are plotted in gray. Figure 7 shows illustrations of all types.

Figure 9 shows the first 3 principal components for the automata, tracked across $\alpha$ and projected onto the $\alpha$-dependent bases. Notice the continuity of the components as the distance is analytic; this procedure is an approximation to functional principal components analysis where the basis vectors are allowed to be functions.

The bottom graph shows the fraction of variance outside the first 1-5 principal components. It is easy to see that as $\alpha \to 0$ the data become 1-dimensional and 1 component suffices. The observed corner in the 3 components line near $\alpha = 0.75$ is due to a shift in the order of eigenvalues, and similarly another shift occurs around $\alpha = 0.85$. The "third" principal component shown here is the one from $\alpha$ close to 1; below 0.75 it is the fourth component. No reordering of the first two components occurs.

The first principal component is the single dimension near $\alpha = 0$, namely initial move, cooperate or defect. Close to $\alpha = 1$ we can see that ALLC and ALLD-type strategies are separated while TFT/PSY-types have 0 coordinate - this can be called "cooperativeness" in the sense of bias towards one move or another. The other two components have 0 coordinates near $\alpha = 0$ as expected for 1-dimensional data. The second component as $\alpha \to 1$ separates the pairs of strategies ALLC/ALLD and TFT/PSY - this can be loosely called "responsiveness" to the opponent. The third component zeros ALLC and ALLD while separating TFT and PSY - loosely this is "coordination", whether the response is equal or opposite to the opponent's move.

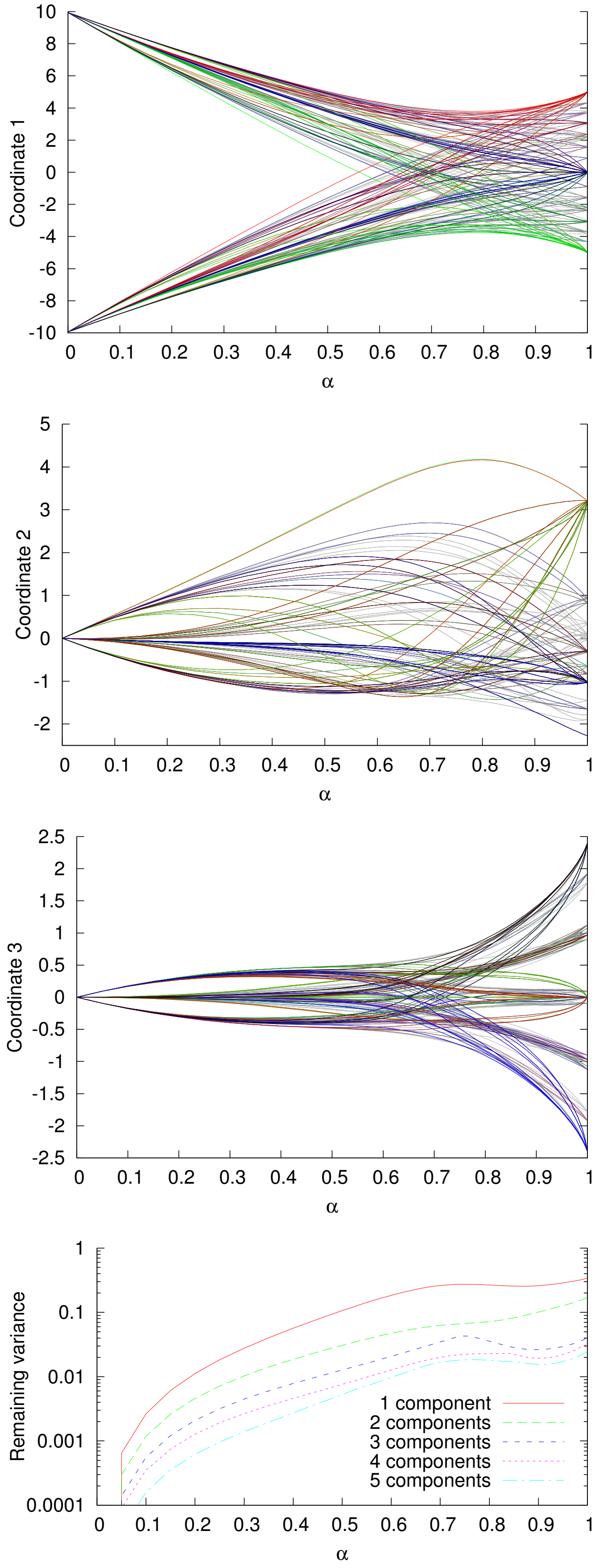

Fig. 9. From top to bottom: the first 3 principal components of all 296 automata tracked over a moving basis on the sample points, and fraction of total variance outside the first 1-5 components.

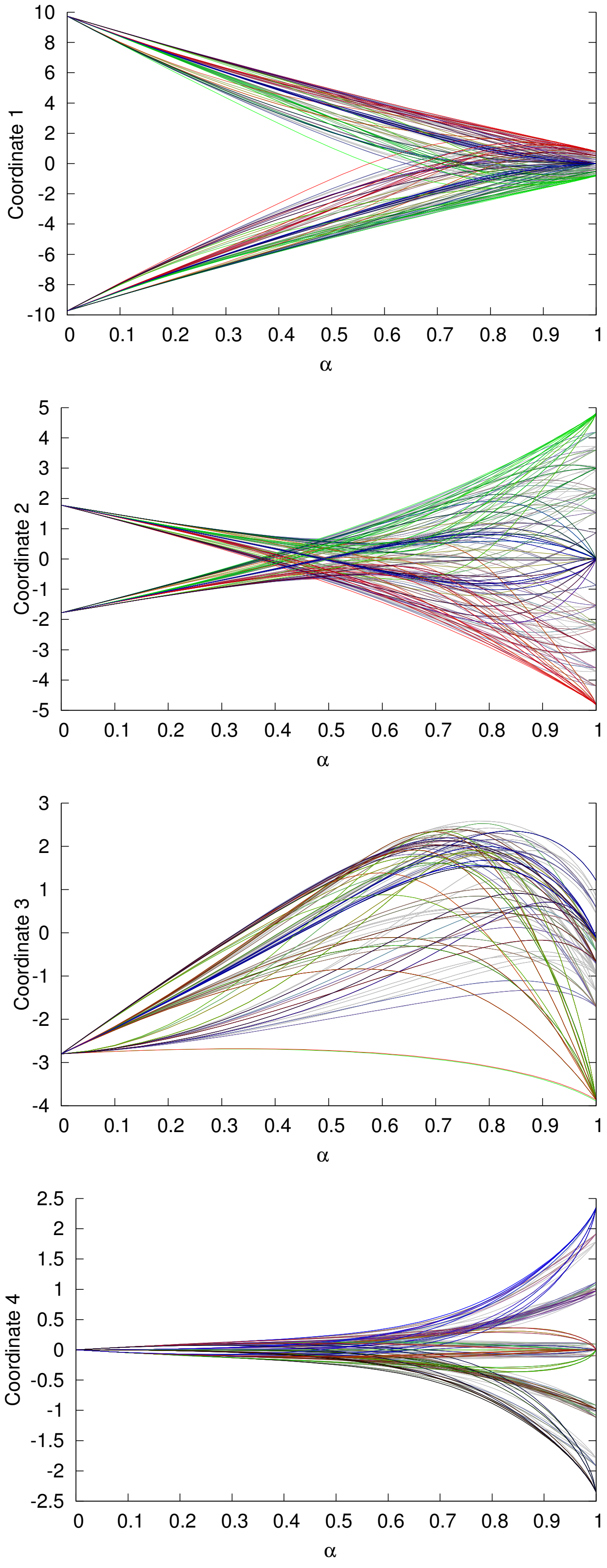

Fig. 10. From top to bottom: the first 4 principal components of all 296 automata tracked over the basis from concatenated data.

Loadings for the various principal components have been omitted since the symmetry of the distance matrices means the loadings are just variance-normalized negatives of the coordinates.

To show the difference between functional PCA and normal PCA with concatenated data, the first 4 coordinates of the automata in concatenated PCA are shown in Figure 10. Notice the similarity between PCA component 1 with the $\alpha = 0$ end of FPCA component 1, PCA component 2 with the $\alpha = 1$ end of same, PCA component 3 with FPCA component 2, and PCA component 4 with FPCA component 3. Hence as $\alpha$ varies, the automata move from a 1-dimensional subspace to a mostly orthogonal 3-dimensional subspace. These 4 components account for 0.958 of the variance.

The 3-dimensional parametric plots of all automata under FPCA basis are shown in Figure 11, along with all 3 planar projections. The array of points as $\alpha \to 1$ can be compared with the $k$-means results at $\alpha = 0.99$: the various points are separated into unequally spaced clusters, and beyond a certain cutoff will split into many subclusters quickly.

## VII. Conclusions and Future Directions

This paper details many aspects of a novel model for fingerprinting game playing agents, including theoretical properties and limitations, computational details and examples of single-player display as well as representation-wide studies. It is shown that finite automata can be fingerprinted in polynomial time in the number of states, and the universal approximability of logically consistent players by automata allows uniform estimation of even non-finite state strategies. Basic analysis of the 2-state deterministic automata reveals that 3 principal components are enough to separate the 8 1-state automata but leave coalesced small subclusters of "similar" strategies in the limit of using the expected value distribution.

Theorem 3 completely partitions the space of move histories into indistinguishable classes, but finding distinct pairs of strategies with identical fingerprints and simple encodings is difficult. Asymptotically the number of degrees of freedom for an arbitrary player outstrips the number of restrictions to achieve functional identity and hence there should be many players that have a given fingerprint. Usual encodings incorporate a finite number of parameters, so it will be interesting to see how many nontrivial pairs exist for some representation. For example, 2-state finite automata have duplicates, but they are all copies of 1-state automata and are hence trivial.

And whilst the distinguishing power of the 1-state probe is easily found, an analogous proof method for $k > 1$ will encounter substantial difficulties. To see this, the response function $\rho_{O_2(\vec{v})}(z, w)$ can be shown to be the solution of the recurrence relations

$$q(i+1) = s_{11w_i} q(i) + s_{12w_i}\big(1 - q(i)\big),\ q(0) = \sum_{m_2 \in \Gamma_2} r_{\emptyset 1 m_2}$$

$$p(i+1) = r_{1z_iw_i} q(i) + r_{2z_iw_i}\big(1 - q(i)\big),$$

$$p(0) = r_{\emptyset 1 w_0} + r_{\emptyset 1 w_1},\ \rho_{O_2(\vec{v})}(z, w) = \prod_{i=0}^{|z|} p(i)$$

where $s_{ijm_2}$ is the probability of transitioning from state $j$ to state $i$ seeing $m_2$ as input. $q(i)$ is the probability of being in state 1 at move $i$ and $p(i)$ is the probability of playing move $i$ as dictated. This will not be a monomial in the parameters, hence to use the same proof orthogonalization is required, and the indistinguishable classes become linear combinations.

The application presented here is on figuring out timescale (per $\alpha$) dependence of the structure of a representation; this can obviously be done for other classes of strategies, for example in [16], [17]. Other questions that can be posed include the mutational connectivity network of the representation, interrelation of the structure of different representations, change in structure when a score matrix is introduced and so forth. However all these are for a static assessment of a single state space, which must necessarily be small or computation required becomes huge.

Given that one of the original motivations for this work was to automate analysis of evolutionary game theory, all aspects of the field can be re-interpreted through the oft-more convenient lens of functions. One direction is to consider the impact of parameters on the population dynamics. These include representation, algorithm choice, parameters of the algorithm, population size and time (explored in [18]), and changes in the game such as different scoring or noise. A concept usable here is of population statistics: this can be easily extracted from a set of functions, and allow nontrivial quantification of phenomena such as evolutionary velocity (change of average) and population diversity (variance).

## Acknowledgment

The author would like to thank Daniel Ashlock for introducing fingerprinting, Rajesh Pereira for suggestions on real analysis, Colin Lee for an explanation of diffusion characters, Chakra Chennubhotla for help with data analysis and preparation, and the Canada Foundation for Innovation for providing the computational resources used in this work.

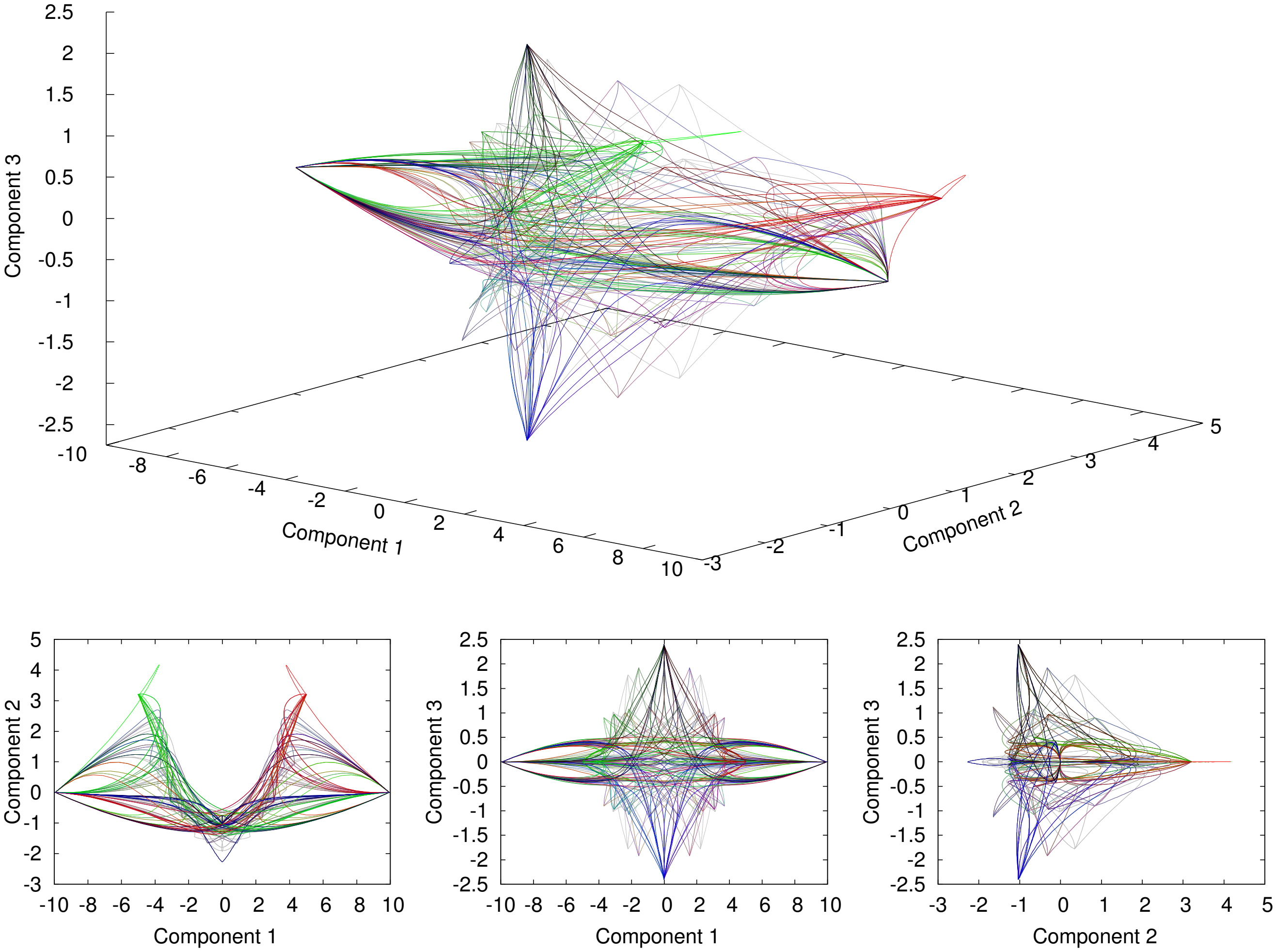

Fig. 11. Line plot of all 296 automata in the first 3 principal component subspace tracked over a moving basis, as well as 3 planar projections. Note that the axes are not on the same scale.

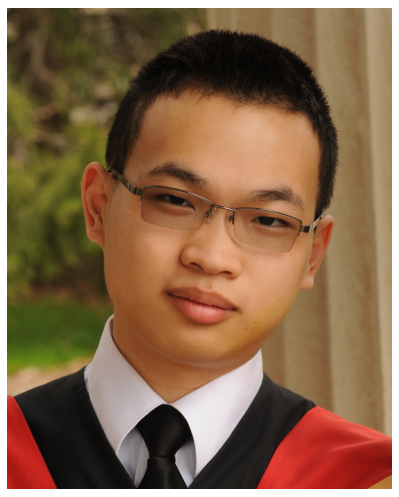

**Jeffrey Tsang** holds a BSc in Mathematics and a Bachelor of Computing from the University of Guelph, ON, Canada, and is currently in the Joint Carnegie Mellon University-University of Pittsburgh PhD Program in Computational Biology, PA, USA.